\documentclass[epj]{svjour}
\usepackage{latexsym,graphicx,epsfig,psfrag,here}    

\def\circa#1{$\!\bigcirc$\hspace{-0.7em}\raisebox{0.05em}
{\scriptsize #1}\hspace{0.2em}}

\begin{document}

\newcommand{\eq}{\begin{eqnarray}}
\newcommand{\en}{\end{eqnarray}}
\renewcommand{\theequation}{\arabic{section}.\arabic{equation}}
\newcommand{\mathbold}[1]{\mbox{\boldmath $\bf#1$}}

\newcommand{\Pint}{-\hspace{-2.3ex}\int}
\newcommand{\bdm}{\begin{displaymath}}
\newcommand{\edm}{\end{displaymath}}
\newcommand{\no}{\nonumber \\}
\newcommand{\sos}{\Delta_{\mbox{\tiny{Roy}}}^2}
\renewcommand{\theequation}{\arabic{section}.\arabic{equation}}
\renewcommand{\Re}{{\rm Re}\,}
\renewcommand{\Im}{{\rm Im}\,}
\newcommand{\be}{\begin{equation}}
\newcommand{\ee}{\end{equation}}
\newcommand{\bea}{\begin{eqnarray}}
\newcommand{\eea}{\end{eqnarray}}
\newcommand{\fs}{\; \; .}
\newcommand{\co}{\; \; ,}

\newcommand{\nn}{\nonumber \\}
\newcommand{\scs}{\co \;}
\newcommand{\sem}{ \; \; ; \;\;}
\newcommand{\per}{ \; .}
\newcommand{\la}{\langle}
\newcommand{\ra}{\rangle}
\newcommand{\bla}{\left\langle}
\newcommand{\unith}{{\bf{\mbox{1}}}}
\newcommand{\MS}{\mtiny{MS}}
\newcommand{\GeV}{\mbox{GeV}}
\newcommand{\MeV}{\mbox{MeV}}
\newcommand{\keV}{\mbox{keV}}
\newcommand{\al}{&\!\!\!\!}
\newcommand{\ind}{\scriptscriptstyle}
\newcommand{\QCD}{\mbox{\scriptsize Q\hspace{-0.1em}CD}}
\newcommand{\qbar}{\overline{\rule[0.42em]{0.4em}{0em}}\hspace{-0.45em}q}
\newcommand{\ubar}{\overline{\rule[0.42em]{0.4em}{0em}}\hspace{-0.5em}u}
\newcommand{\dbar}{\,\overline{\rule[0.65em]{0.4em}{0em}}\hspace{-0.6em}d}
\newcommand{\lbar}{\bar{\ell}}
\newcommand{\bbar}{\bar{b}}
\newcommand{\sbar}{\overline{\rule[0.42em]{0.4em}{0em}}\hspace{-0.5em}s}
\newcommand{\Wbar}{\tilde{W}}
\newcommand{\Pbar}{\tilde{P}}
\newcommand{\Ubar}{\tilde{U}}
\newcommand{\betabar}{\tilde{\beta}}
\newcommand{\lvac}{\langle 0|\,}
\newcommand{\rvac}{\,|0\rangle}
\newcommand{\indR}{\mbox{\scriptsize R}}
\newcommand{\SP}{\hspace{-0.03em}\rule[-0.2em]{0em}{0em}_{\mbox{\tiny\it SP}}}
\newcommand{\hpw}{\hspace{-0.03em}\rule[-0.2em]{0em}{0em}_{\mbox{\tiny\it DF}}}
\newcommand{\as}{\hspace{-0.03em}\rule[-0.2em]{0em}{0em}_{\mbox{\tiny\it H}}}
\newcommand{\B}{\hspace{-0.03em}\rule[-0.2em]{0em}{0em}_{\mbox{\tiny\it B}}}
\newcommand{\cnnnl}{\co\nonumber\\}
\newcommand{\lsim}{\,\raisebox{-0.3em}{$\stackrel{\raisebox{-0.1em}
{$<$}}{\sim}$}\,}
\newcommand{\gsim}{\,\raisebox{-0.3em}{$\stackrel{\raisebox{-0.1em}
{$>$}}{\sim}$}\,}
\newcommand{\sP}{s_{\!\ind P}}
\newcommand{\MP}{M_{\hspace{-0.05em}\ind P}}
\newcommand{\GP}{\Gamma_{\!\ind P}}
\newcommand{\qM}{\kappa}
\newcommand{\Ezero}{\sqrt{\rule[0.1em]{0em}{0.5em}s_0}}
\newcommand{\Eone}{\sqrt{\rule[0.1em]{0em}{0.5em}s_1}}
\newcommand{\Etwo}{\sqrt{\rule[0.1em]{0em}{0.5em}s_2}}

\thispagestyle{empty}

\title{
Ground-state energy of pionic hydrogen to one loop}

\author{J.~Gasser\inst{1}\and M.A.~Ivanov\inst{2}\and E.~Lipartia\inst{3,4,5}\and
M. Moj\v{z}i\v{s}\inst{6,7}\and A.~Rusetsky~\inst{1,5}
}

\institute{{\em{\small Institute for Theoretical Physics, University of
 Bern, Sidlerstrasse 5, 3012 Bern, Switzerland}}
\and
{\em{\small Bogoliubov Laboratory of Theoretical Physics, Joint Institute for 
Nuclear Research, 141980 Dubna, Russia}}
\and
{\em{\small Department of Theoretical Physics 2, Lund University, Solvegatan
 14A, S22362 Lund, Sweden}}
\and
{\em{\small Laboratory of Information Technologies, Joint Institute for
 Nuclear Research, 141980 Dubna, Russia}}
\and
{\em{\small High Energy Physics Institute,Tbilisi State University, University
 St.~9, 380086 Tbilisi, Georgia}}
\and
{\em{\small Department of Physics, University of Massachusetts, Amherst, MA
 01003, USA}}
\and
{\em{\small Department of Theoretical Physics, Comenius University, SK-28415 Bratislava, Slovakia} }}

\abstract{
We investigate the ground-state energy of the $\pi^-p$ atom (pionic
hydrogen) in the framework of QCD+QED. In particular, we evaluate the
strong energy-level shift. We perform the calculation  at
next--to--leading order in the low--energy
expansion in the framework of the relevant effective field theory.
The result provides a relation between the strong energy shift and 
the pion-nucleon $S$-wave scattering lengths 
- evaluated in pure QCD - at next--to--leading order in
isospin breaking and in the low--energy expansion.
We compare our result with available model calculations.
}

\maketitle

\tableofcontents

\setcounter{equation}{0}
\section{Introduction}
\label{sec:intro}

The theory of strong interactions has entered a high 
precision phase in recent years, both experimentally and theoretically.  
On the experimental side, we mention i) the muon 
~anomalous ~magnetic ~moment 
~measurement ~at ~Brookhaven~\cite{gm2}. A calculation of $(g-2)_\mu$ that
matches the foreseen experimental accuracy requires 
that the cross section $e^+e^-\rightarrow \pi^+\pi^-$ in the low--energy 
region is known to better than one percent; ii) experiments that aim
to determine hadronic scattering lengths with high precision are
presently running at CERN~\cite{DIRAC} and at PSI~\cite{PSI-prop,PSI}. 
 On the theory side, the $\pi\pi$ scattering lengths have 
recently been calculated at the few-percent level in 
Ref.~\cite{CGL}. 
In this article, we are concerned with the ongoing experiment on the 
measurement of the energy levels and decays of
the $\pi^-p$ atom (pionic hydrogen) at PSI~\cite{PSI-prop,PSI}. 
It is planned to measure the strong interaction width and shift of the 
ground state at the percent level. These measurements can then be used to 
directly extract from data the $\pi N$ scattering amplitude at threshold.
The aim of the experiment goes, however,  further: it intends to extract 
from these measurements the $S$-wave $\pi N$ scattering lengths 
$a_{0^+}^+ + a_{0^+}^-$ in pure QCD with high precision. In order to
achieve this goal, the relation between the scattering lengths and
the threshold amplitude must be known at an accuracy that matches the
accuracy of the experiment. In other words, one has to remove
isospin--breaking effects from the threshold amplitude with high
precision, in the framework of QCD+QED.

Isospin--breaking effects in low--energy $\pi N$ scattering have been
extensively discussed in the literature on the basis of a potential model
approach. The discussion relevant to the problem of  pionic hydrogen can be
found in Refs.~\cite{Sigg,Rasche-piN}. Here, we rely on the effective 
theory of QCD+QED, a method already invoked~\cite{old-pipi,Bern1,Bern2,Bern4}
in the analogous program for the $\pi^+\pi^-$ atom investigated presently 
at CERN~\cite{DIRAC}. For a critical comparison between the potential model 
and the effective theory framework, we refer the reader to 
Ref.~\cite{potential-PLB}.
  
In the case of pionic hydrogen considered here, the relation between the 
strong energy-level shift $\epsilon_{1s}$ of the ground state and the 
threshold $\pi^-p$ scattering amplitude has been worked out in the effective 
theory in Ref.~\cite{Bern3},
\eq\label{energyshift}
\epsilon_{1s}=-\frac{\alpha^3\mu_c^3
{\cal T}_{\pi N}}{2\pi M_{\pi}}\,
\biggl\{
1-\frac{\alpha(\ln\alpha-1)\mu_c^2{\cal T}_{\pi N}}{2\pi M_{\pi}}\biggr\}
+\cdots\, .\nn
\en
Here, $\mu_c=m_pM_{\pi}(m_p+M_{\pi})^{-1}$ denotes 
the reduced mass\footnote{
We denote the charged pion mass (the proton mass) with $M_\pi (m_p)$.}
of the $\pi^-p$ system,  and ${\cal T}_{\pi N}$ is the threshold amplitude 
for the process $\pi^-p\rightarrow\pi^-p$, evaluated at next--to--leading 
order in isospin breaking. Furthermore, $\alpha\simeq 1/137.036$ denotes 
the fine-structure constant, and the ellipsis stands for the higher-order 
terms in isospin breaking (see below). In the isospin symmetry limit, the 
threshold amplitude is proportional to a particular combination of the 
$S$-wave pion-nucleon scattering lengths $a_{0^+}^\pm$
 (we use the notation of Ref.~\cite{Hoehler}),
\eq\label{sym}
{\cal T}_{\pi N}&=&{\cal T}_{\pi N}^0+\alpha {\cal T}^\gamma
+(m_d-m_u){\cal T}^m,
\nonumber\\[2mm]
{\cal T}_{\pi N}^0&=&4\pi\biggl(1+\frac{M_{\pi}}{m_p}\biggr)\, 
(a^+_{0+}+a^-_{0+})\fs
\en
The ~isospin ~breaking ~amplitudes ${\cal T}^{\gamma,m}$ depend on the 
renormalization group invariant scale of QCD, and on the quark mass 
$\hat{m}=(m_u+m_d)/2$ (we consider two-flavor QCD). Once ${\cal T}^{\gamma,m}$ 
are calculated, a measurement of the shift $\epsilon_{1s}$ allows one 
to determine the combination $a^+_{0+}+a^-_{0+}$. Similarly, a  width 
measurement of the ground state  provides $|a_{0+}^-|$. The values of 
these scattering lengths are correlated e.g. with the pion-nucleon 
sigma-term \cite{sigmaterm}, with the pion-nucleon coupling constant~\cite{GMO},
 and with the Goldberger-Treiman relation, which relates the pion-nucleon 
coupling constant to the axial charge. It goes without saying that the 
scattering lengths are therefore very central objects in the analysis of 
pion-nucleon reactions. We refer the reader to \cite{PSI} for a recent 
investigation of these questions. Here, we concentrate on the evaluation of 
the isospin breaking amplitudes ${\cal T}^{\gamma,m}$.

The amplitudes ${\cal T}^{\gamma,m}$ can 
e.g. be evaluated in ~the ~framework of chiral perturbation theory. 
Writing  the energy shift in the form
\eq\label{LO}
\epsilon_{1s}=-2\alpha^3\mu_c^2\,(a_{0+}^++a_{0+}^-)(1+\delta_\epsilon)\, ,
\en
the leading order calculation~\cite{Bern3} gives 
$\delta_\epsilon=(-4.8\pm2.0)\cdot10^{-2}$, whereas the calculation by 
Ref.~\cite{Sigg} in a potential model framework  leads to  
$\delta_\epsilon=(-2.1\pm0.5)\cdot10^{-2}$.  How should the two calculations 
be compared? As has been pointed out in Ref.~\cite{Bern3}, the leading order 
terms in the effective theory are due to effects that are not all consistently 
taken into account in the potential model calculation. On the other hand, as 
has been emphasized in footnote 1 in Ref.~\cite{PSI}, mass splitting effects in 
strong loops and $\gamma n$ intermediate states show up only at higher orders in 
the chiral expansion. It is the aim of the present article to carry out the 
calculation of  ${\cal T}^\gamma$ and ${\cal T}^m$ at next--to--leading order, 
where these effects come into play.  Further, the interference effect between 
vacuum polarization and strong interactions will be taken into account as 
well\footnote{We are indebted to T. Ericson for pointing out to us that this may be 
an important effect.} - they occur formally at next--to--next--to leading order 
in isospin breaking.

Isospin breaking effects in $\pi N$ scattering have been studied already 
previously [see Ref.~\cite{Fettes} and references therein] by using heavy 
baryon chiral perturbation theory (HBChPT). We were not able to directly use 
these results, for the following reason. In Ref.~\cite{Fettes}, the authors 
calculate physical amplitudes in different channels, and study combinations 
thereof that vanish in the isospin symmetry limit. This is not what is needed
for pionic hydrogen, where one has to extract isospin--breaking correction
to a single amplitude ($\pi^-p$ elastic amplitude, in the case of the energy
levels).  Ref.~\cite{Fettes} does not provide  explicit expressions for the
physical amplitudes, and we  have therefore performed an independent calculation.

In Ref.~\cite{Fettes}, HBChPT was used to calculate the amplitudes. Here, we 
rely on the framework developed by Becher ~and ~Leutwyler~\cite{BL1}. ~This 
method is manifestly Lorentz invariant and preserves chiral power counting 
in the case where baryons and Goldstone bosons are running in the loops. 
In the present context, contributions from virtual photons are needed as well.
We show that the method proposed in Ref.~\cite{BL1} can straightforwardly 
be adapted to this case. Individual Feynman diagrams contain ultraviolet 
(infrared) singularities, due to integration over large (small) momenta. 
We use dimensional regularization to tame both of these singularities. 
The counterterms from the higher order chiral Lagrangians cancel the 
ultraviolet poles at $d=4$  in the final result. In order to check this 
cancellation, we evaluate the 1-loop divergences also with the heat-kernel 
method. Finally, infrared divergences disappear in physical quantities at 
the end of the calculation. In our case, this concerns the 
$\pi^- p\rightarrow\pi^- p$ scattering amplitude at threshold, with the Coulomb 
singularity removed. Lack of phase space forbids the emission of soft photons, 
as a result of which the elastic scattering amplitude must be infrared finite 
at threshold. Needless to say that this cancellation serves as another welcome 
check on our calculation.

The final result for $\delta_\epsilon$ contains, at the next--to--leading order 
~considered ~here, ~several low--energy constants (LECs) that parameterize the 
structure of the effective theory at this order. All but one of these LECs are 
under experimental control or expected to generate a small effect. The remaining 
one, $f_1$, enters the expression already at leading order~\cite{Bern3}. Whereas 
this constant is expected to contribute sizeably to $\delta_\epsilon$, no 
precise determination in terms of experimental data is available yet. We shall 
explain why, in our opinion, a precise determination of $a_{0+}^++a_{0+}^-$ 
from a measurement of the energy-level shift $\epsilon_{1s}$ has to await a
corresponding precise determination of $f_1$, and more importantly, why 
potential models are not of help in this respect.

The layout of the paper is as follows. In section~\ref{sec:kinematics} we 
introduce the definition of the threshold scattering amplitude in the presence 
of photons, and give the relation of this quantity to the strong energy shift 
of the $\pi^-p$ atom. In section~\ref{sec:Lagrangians} we display the set of 
the meson and meson--nucleon chiral Lagrangians, which are used in the 
calculations of the threshold scattering amplitude at $O(p^3)$. In 
section~\ref{sec:tree} we calculate the tree--level contributions to the 
threshold amplitude at $O(p^3)$. In sections~\ref{sec:twopoint_pi} and 
\ref{sec:twopoint_N}, we outline the generalization of the infrared 
regularization procedure~\cite{BL1}, needed in the presence of virtual photons. 
The general procedure for the calculation of the $S$-matrix elements in the
infrared regularization in the presence of virtual photons is described in
section~\ref{sec:smatrix}, where we also discuss the separation of infrared
and ultraviolet divergences. Section~\ref{sec:triangle} deals with the
calculation of the vertex diagram with virtual photons. Here, we
demonstrate the origin of the Coulomb phase and the singular piece of
the scattering amplitude at threshold. In section~\ref{sec:results} we present 
the final result of our calculations of the threshold amplitude, and in 
section~\ref{sec:LECs} we discuss the size of the relevant low--energy constants.
In section~\ref{sec:energyshift}, we evaluate the ground-state energy-level 
shift numerically and compare the result with model calculations. 
Section~\ref{sec:concl} contains our summary and conclusions. Technical 
details are relegated to the appendices.

\setcounter{equation}{0}
\section{Threshold amplitude and the strong energy-level shift}
\label{sec:kinematics}

It is both conventional and convenient to split the ground-state energy into  
electromagnetic and strong parts,
\eq\label{em-s}
E_{1s}=E^{\rm em}_{1s}+\epsilon_{1s}\, .
\en
Despite the fact that this splitting cannot be understood literally
(e.g. there are contributions from strong interactions in $E^{\rm em}_{1s}$), 
it turns out to be useful in the theoretical analysis of hadronic atom data.

The electromagnetic part $E_{1s}^{\rm em}$ has been worked out in 
Refs.~\cite{Sigg,Bern3}. The results of these two investigations - that were
performed in completely different settings - agree numerically with high 
precision. The complete analytic result of the electromagnetic energy at
$O(\alpha^4)$, which is given in Ref.~\cite{Bern3}, includes relativistic
corrections, finite-size effects from proton and pion charge radii, and the
correction due to the anomalous magnetic moment of the proton. In addition,
it includes the contribution from the electron vacuum polarization, which 
arises formally at $O(\alpha^5)$, but is amplified by a large coefficient 
$(M_{\pi}/m_e)^2$. The evaluation of $E_{1s}^{\rm em}$ in Ref.~\cite{Sigg} 
includes in addition some higher-order corrections (vertex corrections, 
corrections to the vacuum polarization contribution), that were omitted in 
Ref.~\cite{Bern3}. Numerically, however, they are much smaller than the ones 
included in both approaches. 

What is exactly measured in the experiment at PSI, is the transition energy 
$E^{\rm meas}_{3p-1s}$ between the $3p$ and $1s$ states~\cite{PSI-prop,PSI}. 
Since the electromagnetic shifts of both levels are known to a very high 
precision~\cite{Sigg}, and since the strong shift in the $3p$ state is very 
small, the measurement allows one to directly extract the strong shift of the 
$1s$ level according to
\eq\label{psi}
\epsilon_{1s}=E^{\rm em}_{3p-1s}-E^{\rm meas}_{3p-1s}\, .
\en
The aim of the new experiment at PSI is to measure $\epsilon_{1s}$ at the 
percent level~\cite{PSI-prop}. Once this measurement is performed, one is 
faced with the task of extracting the strong $S$-wave $\pi N$ scattering lengths
$a_{0+}^++a_{0+}^-$ from this quantity. 

In ~order ~to ~formulate ~the problem rigorously, it is convenient to introduce 
a common counting rule for the isospin--breaking effects, parameterized by the 
fine structure constant $\alpha$ and the up and down quark mass difference 
$m_d-m_u$. We have found it useful to count these effects at the same order, 
introducing the formal parameter $\delta\sim\alpha\sim m_d-m_u$. 
Equation~(\ref{energyshift}), which relates the strong shift $\epsilon_{1s}$ 
to the threshold amplitude, is then valid~\cite{Bern3} up to and including 
terms of order $\delta^4$ - hereafter, this is referred to as the 
``next--to--leading order in isospin breaking''. In the following, it is useful 
to introduce the notation
\eq\label{spl}
{\cal T}_{\pi N}&=&{\cal T}_{\pi N}^0+\delta{\cal T}\, ,\nonumber\\[2mm]
\delta {\cal T}&=&\alpha{\cal T}^\gamma+(m_d-m_u){\cal T}^m\, .
\en 
The ~isospin-symmetric ~part ${\cal T}_{\pi N}^0$ refers to pure QCD, where the 
pion and the nucleon masses are equal, by convention, to the charged pion and 
to the proton masses, respectively. From Eqs.~(\ref{energyshift}), (\ref{sym}) 
and (\ref{LO}), the expression for the isospin--breaking correction 
$\delta_\epsilon$ can now be readily worked out. One still has to add the 
correction $\delta_\epsilon^{\rm vac}$ due to the interference of vacuum
polarization and strong interactions. The complete expression takes the form
\eq\label{deltaE}
\delta_\epsilon&=&
\frac{\delta{\cal T}}{4\pi(1+M_{\pi}/m_p)(a_{0+}^++a_{0+}^-)}
+K+\delta_\epsilon^{\rm vac}\, ,
\nonumber\\[2mm]
K&=&-2\alpha(\ln\alpha-1)\mu_c(a_{0+}^++a_{0+}^-)\, .
\en
The only unknown ingredient in Eq.~(\ref{deltaE}) is thus the isospin--breaking 
part $\delta{\cal T}$ of the threshold amplitude. The chiral expansion of the 
amplitudes ${\cal T}^\gamma$ and ${\cal T}^m$ is performed in the variable 
$r=M_\pi/m_p$ - the expansion of $\delta{\cal T}$ therefore starts at order 
$p^2$ (counting $\alpha$ and $m_d-m_u$ as $O(p^2)$, as usual),
\eq
\delta{\cal T}=\delta{\cal T}_2+\delta{\cal T}_3+O(p^4)\, .
\en
The leading term $\delta{\cal T}_2$ has been determined in Ref.~\cite{Bern3}.
Here, we evaluate $\delta{\cal T}_3$, which amounts to a one--loop
calculation of the $\pi^-p\to\pi^-p$ amplitude. We will show below,
that the chiral expansion of ${\cal T}^m$ starts at $O(r^2)$. At the
accuracy we are concerned with here, the quark mass difference does
thus not enter. [The effect coming from the insertion of the quark 
mass difference into loops may be non-negligible~\cite{Steininger}. 
Its evaluation requires the calculation of the threshold amplitude at 
order $p^4$, which is beyond the scope of this work.]

The contribution $\delta_\epsilon^{\rm vac}$ has been omitted in 
Ref.~\cite{Bern3}, since formally it is higher order in the parameter 
$\delta$. However, numerically it is not negligible, because of the large 
coefficient containing $M_{\pi}/m_e$. The calculation of this correction 
within a non-relativistic effective Lagrangian approach has been carried 
out in Ref.~\cite{Eiras}. Here, we use this result, contained in table II of
Ref.~\cite{Eiras},
\eq\label{delta-vac}
\delta_\epsilon^{\rm vac}=2\,\frac{\delta\Psi(0)}{\Psi(0)}=0.48~\%\, \co
\en
where $\Psi(0)$ stands for the pionic hydrogen wave function at the origin, 
and where $\delta\Psi(0)$ denotes the correction to this wave function due to 
vacuum polarization. The result (\ref{delta-vac}) agrees perfectly with the 
value $\delta_\epsilon^{\rm vac}=0.46~\%$ obtained within the potential model 
of Ref.~\cite{Sigg}.

We now present the exact definition of the threshold amplitude 
${\cal T}_{\pi N}$ which enters the expression~(\ref{energyshift}) for the 
strong energy shift of the ground state. Let us consider the elastic scattering 
process $\pi^-(q)+p(p)\rightarrow\pi^-(q')+p(p')$ in the vicinity of the physical
threshold, at first order in the fine structure constant $\alpha$. External 
momenta are on the mass shell $p^2={p'}^2=m_p^2$, $q^2={q'}^2=M_{\pi}^2$. 
The Mandelstam variables are defined in the standard manner, 
$s=(p+q)^2=(p'+q')^2$, $t=(p-p')^2=(q-q')^2$, $u=(p-q')^2=(p'-q)^2$,
$s+t+u=2(m_p^2+M_{\pi}^2)$. The  3-momentum of the pion and of the nucleon in 
the CM frame is given by 
$|{\bf p}|=\lambda^{1/2}(s,m_p^2,M_{\pi}^2)/(2\sqrt{s})$,
where $\lambda(x,y,z) =(x-y-z)^2-4yz$ denotes the triangle function.
The scattering angle is given by $\cos\theta=1+t/(2{\bf p}^2)$.

\begin{figure}[H]
\begin{center}
\resizebox{0.45\textwidth}{!}{\includegraphics[]{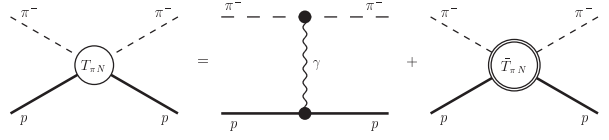}}
\end{center}
\caption{Definition of the truncated $\pi^-p\rightarrow\pi^-p$ amplitude.
The filled vertexes in the diagram with the photon exchange 
correspond to vector form  factors of the pion and of the  proton, 
calculated in pure QCD. 
$\bar T_{\pi N}$ corresponds to the truncated amplitude.} 
\label{fig:1photon}
\end{figure}

The scattering amplitude for this process can be expressed in terms of two
scalar functions $D$ and $B$,\footnote{It is well known that the above 
scattering amplitude in the presence of electromagnetic interactions is 
infrared-singular in perturbation theory. In this paper we use
dimensional regularization to tame both, infrared (IR) and ultraviolet
(UV) divergences. Thus, the scattering amplitudes $D$ and $B$ in
Eq.~(\ref{AB}) are meant to be evaluated at  $d\neq 4$ (see below). We use 
the notation of Bjorken-Drell~\cite{Bjorken} for the Dirac matrices.} 
\eq\label{AB}
T_{\pi N}&=&\bar u(p')\{ D(s,t)
-\frac{1}{4m_p}\,[\not\! q',\!\not\! q]B(s,t)\} u(p)\, ,
\nonumber\\[2mm]
\bar u(p)u(p)&=&2m_p\, .
\en
In the context of the energy-shift considered here, only the amplitude 
$D(s,t)$ is relevant. In the next step, we define the truncated amplitude 
$\bar{D}$, which is obtained from the scattering amplitude (\ref{AB}) by 
subtracting the one-photon exchange contribution displayed in 
Fig.~\ref{fig:1photon},
\eq\label{D}
\bar D(|{\bf p}|,\cos\theta) \doteq \bar D(s,t)
=D(s,t)
-\frac{e^2F_\pi(t)F_1(t)(s-u)}{2m_pt}\, ,\nonumber\\
\en
where $F_\pi(t)$ and $F_1(t)$ denote the pion electromagnetic form factor and 
the nucleon Dirac form factor, respectively.

In the presence of virtual photons, at order $\alpha$, the scattering amplitude 
$\bar D(|{\bf p}|,\cos\theta)$ contains Coulomb singularities at the physical 
threshold~\cite{Bern3,Bern4}. In particular, there emerges the Coulomb phase 
which is divergent in physical dimensions $d=4$ (cf with~\cite{Yennie}), 
\bea\label{theta_C} 
\theta_C(|{\bf p}|)&=&\frac{\mu_c}{|{\bf p}|}\,\mu^{d-4}\, \biggl\{
\frac{1}{d-4}
\nonumber\\[2mm]
&-&\frac{1}{2}\,(\Gamma'(1) 
+\ln4\pi)+\ln\frac{2|{\bf p}|}{\mu}
\biggr\}\fs 
\eea
Here $\mu$ stands for the scale in dimensional regularization.

After removing the Coulomb phase, the behavior of the real part of the
amplitude $\bar D(|{\bf p}|,\cos\theta)$ in the vicinity of threshold, at order
$\alpha$, is given by~\cite{Bern3} 
\eq\label{thr}
&&{\rm Re}\,
\bigl\{{\rm e}^{-2i\alpha\theta_C(|{\bf p}|)}\bar D(|{\bf p}|,\cos\theta)
\bigr\}
\biggr|_{|{\bf p}\to 0}\nonumber\\[2mm] 
&=&\frac{B_1}{|{\bf p}|}+B_2\ln\frac{|{\bf p}|}{\mu_c}
+{\cal T}_{\pi N}+O(|{\bf p}|)\, ,
\en
where the quantities $B_1,~B_2$ can be related to the $S$-wave $\pi N$
scattering lengths (their explicit form is not needed in the applications to
the hadronic atom problem). The quantity ${\cal T}_{\pi N}$,  which is referred 
to in this article as the {\it threshold scattering amplitude}, is infrared 
finite.  It determines $\epsilon_{1s}$ according to Eq.~(\ref{energyshift}),
and is the central object of our investigation.

\setcounter{equation}{0}
\section{Lagrangians}
\label{sec:Lagrangians}

In this section, we display the low--energy effective Lagrangians which are
used in the calculation of the $\pi^-p$ amplitude. In particular, we provide
the meson--nucleon Lagrangian at order $e^2p$, whose divergences are
consistent with the ``standard'' choice of the meson $O(p^4)$~\cite{GL},
$O(e^2p^2)$~\cite{KU}, and nucleon $O(p^3)$~\cite{I,EM} Lagrangians.
The reason for doing so is that the $O(e^2p)$ Lagrangian which is available
in the literature~\cite{Steininger,Muller,thesis}, is not consistent with 
this choice. The details of the derivation, which was performed by using the
Berezinian approach~\cite{Berezinian,Neufeld}, can be found in
appendix~\ref{app:Berezinian}. In this appendix, we in addition give the 
relation between two sets of the LECs which are defined by the choice of the 
$O(p^4)$ meson Lagrangian either in form of Ref.~\cite{GL}, or Ref.~\cite{GSS}.

The full Lagrangian consists of meson and nucleon parts, as well as the free 
photon Lagrangian together with the gauge-fixing term,
\eq\label{L_eff}
{\cal L}_{eff}&=&{\cal L}_\pi+{\cal L}_N+{\cal L}_\gamma\, ,
\nonumber\\[2mm]
{\cal L}_\pi&=&{\cal L}_\pi^{(p^2)}+{\cal L}_\pi^{(e^2)}
+{\cal L}_\pi^{(p^4)}
+{\cal L}_\pi^{(e^2p^2)}+\cdots\, ,
\nonumber\\[2mm]
{\cal L}_N&=&{\cal L}_N^{(p)}+{\cal L}_N^{(p^2)}
+{\cal L}_N^{(e^2)}
+{\cal L}_N^{(p^3)}+{\cal L}_N^{(e^2p)}+\cdots\, ,\nonumber\\
\en
where
\eq\label{L_2}
&&{\cal L}_\pi^{(p^2)}+{\cal L}_\pi^{(e^2)}+{\cal L}_\gamma
=\frac{F^2}{4}\,\langle d^\mu U^\dagger d_\mu U
+\chi^\dagger U+U^\dagger\chi\rangle
\nonumber\\[2mm]
&&+ZF^4\langle {\cal Q}U{\cal Q}U^\dagger\rangle
-\frac{1}{4}\,F_{\mu\nu}F^{\mu\nu}
-\frac{1}{2a}(\partial_\mu A^\mu)^2\, ,
\\[5mm]\label{L_x}
&&{\cal L}_\pi^{(p^4)}=\sum_{i=1}^{7}l_iO_i^{(p^4)}\, ,
\qquad
{\cal L}_\pi^{(e^2p^2)}=F^2\sum_{i=1}^{10}k_iO_i^{(e^2p^2)}\, ,
\nonumber\\[2mm]
&&{\cal L}_N^{(p)}=\bar\Psi(i\not\!\! D-m
+\frac{1}{2}\, g_A\not\! u\gamma_5)\Psi\, ,
\nonumber\\[2mm]
&&{\cal L}_N^{(p^2)}=\sum_{i=1}^7 c_i \bar\Psi\, O_{i}^{(p^2)}\Psi\, ,
\qquad\!
{\cal L}_N^{(e^2)}=F^2\sum_{i=1}^3 f_i \bar\Psi\, O_{i}^{(e^2)}\Psi\, ,
\nonumber\\[2mm]
\label{L_p3}
&&{\cal L}_N^{(p^3)}=\sum_{i=1}^{23}d_i\bar\Psi O_i^{(p^3)}\Psi\, ,
\qquad\!\!
{\cal L}_N^{(e^2p)}=F^2\sum_{i=1}^{12}g_i\bar\Psi O_i^{(e^2p)}\Psi\, ,
\nonumber\\
&&
\en
where $\langle A\rangle$ denotes the trace of the matrix $A$.
The building blocks for the mesonic Lagrangians are
\eq\label{derivative_pi}
&&d_\mu U=\partial_\mu U-i{\cal R}_\mu U+iU{\cal L}_\mu\, ,\quad
\nonumber\\[2mm]
&&
\pmatrix{{\cal R}_\mu\cr {\cal L}_\mu}=v_\mu\pm a_\mu+{\cal Q}A_\mu\, ,
\nonumber\\[2mm]
&&{\cal R}_{\mu\nu}=\partial_\mu {\cal R}_\nu-\partial_\nu 
{\cal R}_\mu-i[{\cal R}_\mu,{\cal R}_\nu]\, ,\quad
\nonumber\\[2mm]
&&{\cal L}_{\mu\nu}=\partial_\mu {\cal L}_\nu-\partial_\nu 
{\cal L}_\mu-i[{\cal L}_\mu,{\cal L}_\nu]\, ,
\nonumber\\[2mm]
&&{\cal F}_\pm^{\mu\nu}=u^\dagger{\cal R}^{\mu\nu}u\pm 
u{\cal L}^{\mu\nu}u^\dagger\, ,
\nonumber\\[2mm]
&&\hat {\cal F}_\pm^{\mu\nu}={\cal F}_\pm^{\mu\nu}-\frac{1}{2}\,
\langle {\cal F}_\pm^{\mu\nu}\rangle\, ,
\nonumber\\[2mm]
&&\chi = 2 B (s+ip)\, ,\quad
d_\mu \chi=\partial_\mu \chi-i{\cal R}_\mu \chi+i\chi{\cal L}_\mu\, ,
\nonumber\\[2mm]
&&{\cal Q}_\pm=\frac{1}{2}\,(u{\cal Q}u^\dagger\pm u^\dagger{\cal Q}u)\, ,
\nonumber\\[2mm]
&&{\cal C}_R^\mu=-i[{\cal R}_\mu,{\cal Q}]\, ,\quad
{\cal C}_L^\mu=-i[{\cal L}_\mu,{\cal Q}]\, ,\quad
\nonumber\\[2mm]
&&{\cal C}_\pm^\mu=\frac{1}{2}\,(u {\cal C}_L^\mu u^\dagger
\pm u^\dagger {\cal C}_R^\mu u)\, ,
\en
where $U$ is a unitary $2\times 2$ matrix, and $\Psi$ is the nucleon field. 
As usual, $s,p,v_\mu,a_\mu$ denote external scalar, pseudoscalar, vector and 
axial fields, $B$ is a constant related to the quark condensate, and
${\cal Q}=e\,{\rm diag}(\frac{2}{3},-\frac{1}{3})$ is the quark charge 
matrix. The axial source is taken traceless, $\langle a_\mu\rangle=0$.

Building blocks for the meson--nucleon Lagrangians in the presence of virtual 
photons are
\eq\label{derivative}
&&D_\mu=\partial_\mu+\Gamma_\mu\, ,\quad
U=u^2\, ,
\quad u_\mu=iu^\dagger d_\mu Uu^\dagger\, ,
\nonumber\\[2mm]
&&\Gamma_\mu=\frac{1}{2}\,[u^\dagger,\partial_\mu u]
-\frac{i}{2}\, u^\dagger R_\mu u
-\frac{i}{2}\, u L_\mu u^\dagger\, ,
\nonumber\\[2mm]
&&\chi_\pm=u^\dagger\chi u^\dagger\pm u\chi^\dagger u\, ,\quad
\hat\chi_+=\chi_+-\frac{1}{2}\,\langle\chi_+\rangle\, ,
\nonumber\\[2mm]
&&F^\pm_{\mu\nu}=u^\dagger R_{\mu\nu}u\pm uL_{\mu\nu}u^\dagger\, ,\quad
\hat F^+_{\mu\nu}=F^+_{\mu\nu}-\frac{1}{2}\,\langle F^+_{\mu\nu}\rangle\, ,
\nonumber\\[2mm]
&&Q_\pm=\frac{1}{2}\,(uQu^\dagger\pm u^\dagger Qu)\, ,\quad
\hat Q_\pm=Q_\pm-\frac{1}{2}\,\langle Q_\pm\rangle\, ,
\nonumber\\[2mm]
&&c^\pm_\mu=-\frac{i}{2}\,(u[L_\mu,Q]u^\dagger\pm u^\dagger[R_\mu,Q]u)\, ,
\en
where $Q=e\,{\rm diag}(1,0)$ denotes the nucleon charge matrix, 
and $R_\mu$, $L_\mu$, $R_{\mu \nu}$, $L_{\mu \nu}$ are defined just like their 
pionic counterparts ${\cal R}_\mu$, ${\cal L}_\mu$, ${\cal R}_{\mu \nu}$, 
${\cal L}_{\mu \nu}$ respectively, with ${\cal Q}$ replaced by $Q$. We have set 
the charge matrices ${\cal Q}$ and $Q$ to their constant physical values, and do
not consider the most general expression of the effective Lagrangians
containing space-dependent spurion fields $Q_{L,R}$. Furthermore,  in
(\ref{L_x}) we drop terms which do not contain pion fields, and terms of 
order $e^4$.

We comment on the LECs $F,g_A,Z,c_i,f_i,\cdots$. The first one, $F$,
denotes the pion decay constant in the chiral limit, and $g_A$ is the axial 
charge, again at $m_u=m_d=0$. The quantity $Z$ is related to the pion mass 
difference,
\eq\label{pion-Z}
\hspace*{-10.mm}&&
M_{\pi^0}^2=(m_u+m_d)B
+O(m_q^2,e^2m_q,e^4)\co\nonumber\\[2mm]
\hspace*{-10mm}
&&\Delta_\pi=M_{\pi}^2-M_{\pi^0}^2
=2e^2F^2Z+O(m_q^2,e^2m_q,e^4) .
\en
The couplings $c_i,~f_i$ are finite, if the calculation of the loop diagrams 
is performed in a manner that respects chiral power counting. The divergences 
in the remaining LECs are given by

\eq\label{div_meson_nucleon}
\hspace*{-0.3cm}
l_i&=&\gamma_i\lambda+l_i^r(\mu)\, ,\quad
k_i=\sigma_i\lambda+k_i^r(\mu)\co\nonumber\\[2mm]
d_i&=&\frac{\beta_i}{F^2}\,\lambda+d_i^r(\mu)\, ,\quad
g_i=\frac{\eta_i}{F^2}\,\lambda+g_i^r(\mu)\, ,
\en
where
\eq\label{div}
\lambda=\frac{\mu^{d-4}}{16\pi^2}\,\biggl(\frac{1}{d-4}-\frac{1}{2}
[\Gamma'(1)+\ln 4\pi+1]\biggr)\, .
\en

In appendix~\ref{app:lagrtables} we list the operators $O_i^{(k)}$, as well as 
the divergent parts of the low--energy constants, and provide references to the 
original literature. Since our operator basis in the mesonic sector is defined 
in a standard manner, numerical values of $l_i^r,~k_i^r$ can be directly taken 
from the existing analyses. The definition of the finite constants $c_i$ and 
$f_i$ coincides with that from Refs.~\cite{I} and ~\cite{Muller}, respectively 
(in the latter, the notation $f'_i$ instead of $f_i$ is used for the LECs in the
relativistic theory). Our choice for those $d_i^r$ that contribute to the 
$\pi N$ scattering amplitude, corresponds to the one of Ref.~\cite{BL2}, modulo 
the relations~(\ref{d5new}), which display the effect of the different choice 
of the mesonic basis here and in Ref.~\cite{BL2}. Finally, for the reasons 
given in  appendix~\ref{app:Berezinian}, it is not clear to us how we can 
compare the couplings $g_i^r$ used here, with the ones determined in 
Ref.~\cite{Fettes}.

In the actual calculation of the scattering amplitude, we invoke an exponential 
parameterization of the matrix $U$,
\eq\label{exponential}
&&U=\exp{i\mathbold{\pi}}/{F}\, ,\qquad
\mathbold{\pi}=\pmatrix{\pi^0 & \sqrt{2}\pi^+\cr \sqrt{2}\pi^- & -\pi^0}\, ,
\en
and use $\mathbold{\pi}$ as an interpolating field for the pions. Further, we
set 
\eq
\chi=2B\, {\rm diag}(m_u,m_d)\, ,
\en
and perform the calculations in the Feynman gauge $a=1$.

\setcounter{equation}{0}
\section{Tree contributions}
\label{sec:tree}
\begin{figure}[H]
\begin{center}
\resizebox{0.35\textwidth}{!}{\includegraphics*{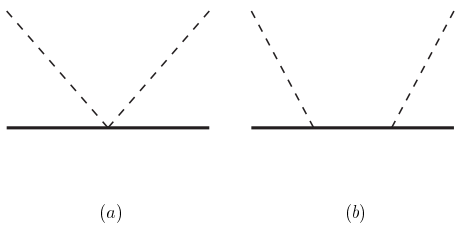}}
\end{center}
\caption{The Born diagrams, vector (a) and axialvector (b) couplings. }
\label{fig:Born}
\end{figure}

\begin{figure}[H]
\begin{center}
\resizebox{0.35\textwidth}{!}{\includegraphics*{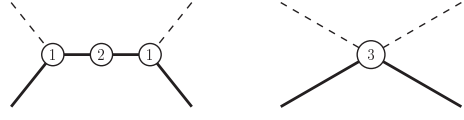}}
\end{center}
\vskip-.5cm
\caption{Counterterm contributions. Insertions on the external lines
are not shown.}
\label{fig:counterterms}
\end{figure}

The tree contributions to the threshold scattering amplitude ${\cal T}_{\pi N}$
include the pseudovector Born diagrams  Fig.~\ref{fig:Born}, and the 
contributions from the counterterms in the effective Lagrangians, indicated in 
Fig~\ref{fig:counterterms}. Insertions in the external lines are not 
displayed - they will be included through mass and wave function renormalization 
in section~\ref{sec:results}.

The scattering matrix elements are evaluated with  meson momenta that are  on 
the mass shell, $q^2=q'^2=M_{\pi}^2$,  and with  spinors that obey the Dirac 
equation with the physical proton mass, $\not\!\! p u(p,s)=m_p u(p,s)$. On the 
other hand, the neutron mass in the internal line in Fig~\ref{fig:Born}b is the 
chiral symmetric one. It receives corrections from counterterm 
insertions~\circa{2}\, in Fig.~\ref{fig:counterterms}, and from  the self energy 
diagrams $s_{1}$ and $s_{2} $ in Fig.~\ref{fig:strong}. Here, we merge these 
contributions and evaluate the diagram Fig~\ref{fig:Born}b with the physical 
neutron mass. [The subtraction in the one--loop self-energy-type diagrams 
$s_{1}$ and $s_{2} $ in Fig.~\ref{fig:strong} is then carried out on the mass 
shell.] In this convention, the contribution from the Born diagrams 
Fig.\ref{fig:Born} becomes
\eq\label{Born}
{\cal T}^{\rm B}=\frac{M_{\pi}}{2F^2}
-\frac{g_A^2M_{\pi}^2}{2F^2(m_n+m_p+M_{\pi})}\fs
\en
Next  consider the vertex corrections \circa{1}\, -  these can be absorbed in 
$g_A$. Since they are proportional to the quark masses $m_u,m_d$, they do not 
contribute at order $p^3$, and we drop them altogether. The polynomial 
contributions \circa{3}\, arise from  two-nucleon two-pion vertexes in the 
effective Lagrangians. We find\footnote{The normalization of the low--energy 
constants $f_{1,2}$ used in the present paper differs from that of 
Ref.~\cite{Bern3}: $F^2f_{1,2}^{\rm here}=f_{1,2}^{\rm old}$.}
\eq\label{c.t.}
&&{\cal T}^{\rm ct}=-\frac{4M_{\pi^0}^2}{F^2}\, c_1
+\frac{2M_{\pi}^2}{F^2}\,(c_2+c_3)
-\frac{e^2}{2}\,(4f_1+f_2)\nonumber\\[2mm]
&&+\frac{4M_{\pi}^3}{F^2}\,(d_1+d_2+d_3)
+\,\frac{8M_{\pi}M_{\pi^0}^2}{F^2}\, d_5
\nonumber\\[2mm]
&&
+2e^2M_{\pi}(g_6+g_8)\, .
\en

\setcounter{equation}{0}
\section{Two--point functions of pion fields}
\label{sec:twopoint_pi}

In Ref.~\cite{BL1}, Becher and Leutwyler have shown how power counting in  
baryon chiral perturbation theory (pions and nucleons running in the loops) 
can be incorporated in a manifestly Lorentz invariant manner. Here, we extend 
this framework to include virtual photons. In addition to the question of power 
counting, virtual photons generate the  standard problems: poles in two--point 
functions are transformed into branch points,  wave function renormalization 
constants become ill-defined, and truncated on--shell Green functions in general 
cease to exist in four dimensions. In order to identify these infrared 
singularities\footnote{There are two types of infrared singularities in the 
present context: The ones associated with non--analytic terms in the chiral 
expansion, and singularities generated by the presence of photons. Since it will 
always be clear what singularities we have in mind, we do not distinguish in 
the following between the two, which makes notation less clumsy.}, we start the 
discussion of the low--energy expansion of Green functions  with the two--point 
function of pseudoscalar quark currents. In this case, power counting does not 
pose a problem, because one is concerned with mesons and photons only. 
Furthermore, ultraviolet divergences do not show up  in the final result  
either, because Green functions of quark currents are well defined objects in 
the effective Lagrangian framework. Therefore, the only obstacles in this 
case are the infrared divergences generated by photon loops.

\subsection{Pseudoscalar two--point function}
The ~pseudoscalar ~densities 
$P_5^{\pm}(x)=\qbar (x)i\gamma_5\tau^{1\pm i2}q(x)$ 
 may be used as interpolating fields for the charged pions. To evaluate 
scattering matrix elements, the residue of the two--point function
\bea
G(s)=i\int dx e^{-iqx}
\la 0|TP_5^-(x) P_5^+(0)|0\ra\sem
s=q^2 \nonumber\\
\eea
at $s=M_{\pi}^2$ is needed - it determines  the wave function renormalization 
constant. However, in the presence of virtual photons, $G$ develops a 
branchpoint, not a pole, and that residue does not exist.  There are two 
standard procedures to deal with the situation: Either, one introduces a 
photon mass, which shifts the branchpoint to $s=(M_{\pi}+m_\gamma)^2$, as a 
result of which $G$ indeed contains  a pole at the pion mass. The photon mass 
is then sent to zero at the very end of the relevant calculations of physical 
quantities. Or one instead uses dimensional regularization to tame both, 
ultraviolet and infrared divergences. In the following, we use the latter method.
Let us consider the low energy expansion of $G$. Including contributions from 
one virtual photon as indicated in 
\begin{figure}[thp]
\begin{center}
\resizebox{0.3\textwidth}{!}{\includegraphics{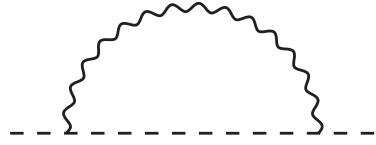}} 

\vspace*{-10.mm}

\caption{Photon contribution to the two--point function of  
pseudoscalar fields.}
\label{fig:pseudoscalar}
\end{center}
\end{figure}
Fig.~\ref{fig:pseudoscalar}, and adding the counterterms from 
 ${\cal L}_\pi^{(e^2p^2)}$ in Eq.~(\ref{L_eff}), we find
\bea
G(s)&=&\frac{4F^2B^2 R(s)}{M_{\pi}^2-s}\co \nonumber\\
M_{\pi}^2&=&M^2-\frac{e^2M^2}{16\pi^2}\left\{3\ln\frac{M^2}{\mu^2}
-4\,\right\}
\nonumber\\[2mm]
&-&e^2M^2K_M^r+\cdots\sem 
\quad\quad
M^2=(m_u+m_d)B\, ,
\nonumber\\
R(s)&=&1-\frac{e^2}{8\pi^2}\left\{r(s)
+\ln{\frac{M^2}{\mu^2}}-1\,\right\}
-e^2K_G^r+\cdots\co
\nn
r(s)&=&\frac{s+M^2}{s}\ln\frac{M^2-s}{M^2}\co\nn\
K_M^r&=&\frac{20}{9}(k_1^r+k_2^r-k_5^r)
-\frac{92}{9}k_6^r
-\frac{4}{9}k_7^r-8k_8^r\co\nn
K_G^r&=&\frac{20}{9}(k_1^r+k_2^r-2k_5^r-2k_6^r)
-\frac{8}{9}k_7^r-8k_8^r\fs
\eea
[Here, we have omitted the contributions from the leading electromagnetic 
term proportional to $Z$ in Eq.~(\ref{pion-Z}), such that 
$M_{\pi}^2=M_{\pi^0}^2$ at leading order.] The ellipses denote additional 
terms in the low--energy expansion. The correlator $G(s)$ is finite at 
nonexceptional momenta, but the residue develops a logarithmic singularity at 
$s=M_{\pi}^2$, manifest in the function $r(s)$ [at the order of the expansion 
considered here, we may replace $M^2$ in the function $r(s)$ by $M_{\pi}^2$].

In order to continue, we keep the dimension different from four until the very 
end of the calculation of physical quantities. The function $r(s)$ becomes
\bea
r(s,d)&=&\frac{a+1}{a-1}(4\pi)^{\omega}
\Gamma(\omega)M^{-2\omega}
\int_0^1dx x^{-\omega}\nonumber\\[2mm]
&\times&\left[x^{-\omega}\right.
\left.
-(1-a+ax)^{-\omega}
\right]
+O(d-4)\co\nonumber\\[2mm]
\omega&=&2-\frac{d}{2}\, ,\qquad a=s/M^2\, .
\eea
At nonexceptional momenta, $r(s,d)$ approaches $r(s)$ as $d\rightarrow 4$. 
If $d$ is bigger than four, one may perform the mass-shell limit,
\bea
R(s,d)&=&
1-\frac{e^2}{8\pi^2}\left[-C_{\rm IR}^\pi +\ln{\frac{M^2}{\mu^2}}\right]
\nonumber\\[2mm]
&-&e^2K_G^r+O(d-4)\sem\, \quad\quad
s \rightarrow M_{\pi}^2\co
\nonumber\\[2mm]
C_{\rm IR}^\pi&=&2M_{\pi}^{d-4}\left(\frac{1}{d-4}\right.
\left.-\frac{1}{2}\,(\Gamma'(1)+\ln(4\pi)+1)\right)\fs\nonumber\\
\eea
The infrared singularity manifests itself in a pole at $d=4$, which  is due to  
integration over small loop momenta. We indicate the origin of this singularity 
with the symbol ${\rm IR}$ in the divergent quantity $C_{\rm IR}^\pi$.

We do not  use pseudoscalar densities as interpolating fields - it is simpler 
to use the pion fields instead. In this case, additional singularities occur 
at $d=4$, as is shown in the following subsection.

\subsection{Two--point function of the pion fields}
The Green functions of the pion fields depend on the parameterization used 
for the matrix $U(x)$ in the effective Lagrangian. Here, we use the exponential 
parameterization (\ref{exponential}), and consider the propagator of the 
charged fields,
\bea\label{eq:twopointpion}
\frac{R_\pi(s,d)}{M_{\pi}^2-s}=
i\int dx e^{-iqx}
\la 0|T\pi^-(x) \pi^+(0)|0\ra\sem
s=q^2\fs\nonumber\\
\eea
We again include the photon loop from Fig.~\ref{fig:pseudoscalar} and the 
pertinent counterterms from ${\cal L}_\pi^{(e^2p^2)}$ at $Z=0$. The result for 
the pion mass is the same as before, whereas  several of the counterterms 
$k_i$ are now absent in the residue. We find
\bea
&&R_\pi(s,d)=1-\frac{e^2}{32\pi^2}\left[4r(s,d) + C_{\rm UV}^\pi
+3\ln{\frac{M^2}{\mu^2}}-4\right]
\nonumber\\[2mm]
&&- e^2K_\pi^r+\cdots\, , \quad\quad K_\pi^r=\frac{20}{9}(k_1^r+k_2^r)\co
\nonumber\\[2mm]
&&C_{\rm UV}^\pi=2M_{\pi}^{d-4}
\left(\frac{1}{d-4}\right.
\left.-\frac{1}{2}\,(\Gamma'(1)+\ln(4\pi)+1)\right)\fs\nn
\eea
In contrast to the two--point function $G(s)$, the propagator of the pion 
field is ultraviolet divergent: the function $R_\pi$ develops a pole in four 
dimensions also at nonexceptional momenta. We indicate the origin of this 
singularity with the symbol UV in the divergent quantity $C_{\rm UV}$. The 
reason for the occurrence of this pole is well understood: the effective 
theory guarantees that Green functions of quark currents are ultraviolet 
finite, whereas the pion fields simply serve as integration variables in the 
path integral and are devoid of any physical significance. 

We conclude that dimensional regularization generates two different types of 
poles at $d=4$: those that are due to integrations over large loop momenta, 
and those that are due to considering the mass shell restriction (integrations 
over small loop momenta). In each graph, one may easily distinguish between the
two types of singularities, and we will do so below, by denoting the
corresponding singular quantities with indices ${\rm UV}$ and ${\rm IR}$,
respectively. As we shall see later, distinguishing between the two divergences 
serves as a powerful check on our calculation.

\begin{figure}[thb]
\begin{center}
\resizebox{0.3\textwidth}{!}{\includegraphics*{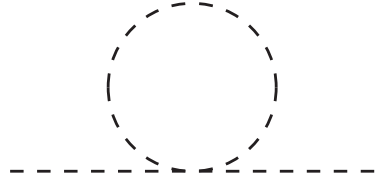}}
\end{center}

\vspace*{-10.mm}

\caption{Pion loop contributions to the two--point function 
(\protect{\ref{eq:twopointpion}}).
Neutral and charged pions are running in the loop.}
\label{fig:pionGF}
\end{figure}

Later in this article, we also need the contributions from the pion loops 
displayed in Fig.~\ref{fig:pionGF}. Including also the terms generated by the 
leading electromagnetic term proportional to $Z$ which contributes to the pion 
mass difference, we find  
\bea\label{Z-pi}
Z_\pi&=& R_\pi(M_{\pi}^2,d)= 
1+\frac{e^2}{8\pi^2}\,(C^\pi_{\,\rm IR}- C^\pi_{\,\rm UV})
\nonumber\\[2mm]
&+&\frac{1}{24\,\pi^2\,F^2}\,(M^2_{\pi}-e^2F^2Z)\,
C^\pi_{\,\rm UV}
-\frac{e^2Z}{24\pi^2}
\nonumber\\[2mm]
&-&\frac{20\,e^2}{9}\,(k_1+k_2) + O(e^2M^2,M^4,e^4)+O(d-4)\fs\nn
\eea
Here, it is convenient to use the unrenormalized couplings $k_{1,2}$.

\setcounter{equation}{0}
\section{Two--point function of baryon fields}
\label{sec:twopoint_N}

\subsection{Self-energy of the heavy scalar field}

\begin{figure}[thb]
\begin{center}
\resizebox{0.3\textwidth}{!}{\includegraphics*{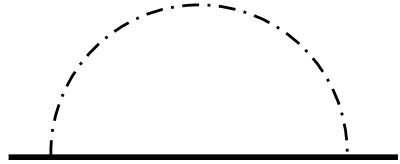}} 
\end{center}
\caption{Self-energy diagram for a heavy field of mass $m$. 
The solid line denotes the heavy field, and the dash-dotted 
line stands for a massless scalar particle.}
\label{fig:heavyfield}
\end{figure}

We now consider the case where a heavy particle interacts with the photon.
To illustrate the method to preserve chiral power counting in this case, 
we consider the self energy diagram Fig.~\ref{fig:heavyfield}, 
\noindent
which represents a heavy scalar particle of mass $m$, interacting with a 
massless field,
\bea\label{eqselfN}
\Sigma(s,d)
&=&\frac{1}{i(2\pi)^d}\int\! \frac{d^dk} 
{[-k^2][m^2-(p-k)^2]}
\nonumber\\[2mm]
&=&(4\pi)^{-(2-w)}\Gamma(w)\int_0^1 dx z^{-w}\, ,
\nonumber\\[2mm]
z&=&x(m^2-s(1-x))\sem s=p^2\, .
\eea 
Evaluation in the standard manner shows that $\Sigma$  is of chiral order zero, 
$\Sigma \sim \mbox{const.}$, near the mass shell, whereas power counting 
suggests that it is of order one, $\Sigma\sim (m^2-s)/m^2$. In Ref.~\cite{BL1}, 
it has been shown that power counting with nucleon and pion fields is preserved,
if one  extends the integration over the  Feynman parameter - that combines the 
heavy and light fields - from zero to infinity.  Proceeding in the same manner 
for photon fields, we find that
\bea
&&\Sigma^I(s,d)
=(4\pi)^{-(2-w)}\Gamma(w)\int_0^\infty dx z^{-w}
\nonumber\\[2mm]
&&=(4\pi)^{-(2-w)}\Gamma(w)\Omega^{1-2w}s^{-w}
\int_0^\infty dx x^{-w}(1+x)^{-w}
\sem
\nonumber\\[2mm]
&& \Omega=(m^2-s)/s\, ,
\eea
where the index $I$ denotes the infrared part a la Becher-Leutwyler. It is seen 
that  the self energy diagram is now of chiral order one near threshold. The 
mass shell constraint is performed by first continuing the result to dimensions 
bigger than four, as before. One finds that
\bea\label{eqsigmaIR}
\Sigma^I(m^2,d)={\Sigma^I}'(m^2,d)=0\scs d>4\co
\eea
where the prime denotes a derivative with respect to $s$. In other words, the 
self energy diagram Fig.\ref{fig:heavyfield} does contribute neither to the 
mass nor to the residue of the two--point function in this prescription.

We found it very useful to  distinguish between infrared and ultraviolet 
divergences also here. This may be achieved as follows. We consider the self 
energy and its derivative at $s=m^2$. From Eq.~(\ref{eqselfN}), one has 
\bea\label{c_uvir}
\Sigma^S(m^2,d)&=&\frac{1}{16\pi^2}(-C_{\,\rm UV}+1)\co\nn
C_{\rm K}&=&2m^{d-4}\left(\frac{1}{d-4}\right.
\left.
-\frac{1}{2}\,(\Gamma'(1)+\ln(4\pi)+1)\right)\co 
\nonumber\\[2mm]
K&=&\rm UV,IR\, .
\eea
We have indicated with the index $S$ the standard procedure, where  the 
integral over the Feynman parameter is performed from zero to one. We have 
booked the singularity at $d=4$ as an ultraviolet one for obvious reasons. 
The derivative is
\bea
{\Sigma^S}'(m^2,d)&=&\frac{1}{32\pi^2m^2}\,(C_{\,\rm IR}-1)\co
\eea 
where the singularity at $d=4$ is now due to integrations over small momenta 
and therefore booked as an infrared singularity. The Becher-Leutwyler infrared 
regularized part is obtained by  subtracting from these expressions the  
'regular part' of the Feynman diagram,
\bea
\Sigma^R(s,d)=-(4\pi)^{-(2-w)}\Gamma(w)
\int_1^\infty dx z^{-w}\fs
\eea
This integral converges for $d < 3$ - the relevant poles at $d=4$ are therefore 
booked as ultraviolet. We find
\bea
\hspace*{-7.mm}&& \hspace*{-3.mm}
\Sigma^R(m^2,d)=\frac{1}{16\pi^2}
(-C_{\,\rm UV}+1)\co\nn
\hspace*{-7.mm}&&\hspace*{-3.mm}{\Sigma^R}'(m^2,d)=
\frac{1}{32\pi^2 m^2}(C_{\,\rm UV}-1)\co
\eea
and therefore
\bea
\Sigma^I(m^2,d)
&=&\Sigma^S(m^2,d)-\Sigma^R(m^2,d)=0\, ,
\\[2mm]
{\Sigma^I}'(m^2,d)
&=&\frac{1}{32\pi^2m^2}(C_{\,\rm IR}-C_{\,\rm UV})\,. 
\eea
The result for the value of the self energy on the mass shell agrees with 
Eq.~(\ref{eqsigmaIR}), whereas its derivative becomes the same upon identifying 
$C_{\, \rm IR}$ with $C_{\, \rm UV}$\, .

\subsection{The nucleon propagator}
The infrared regularized nucleon propagator is evaluated in a completely 
analogous manner, applying infrared regularization to both, the pion and the 
photon  field. In the following, we only need the wave function renormalization 
constant for the proton field. We define it in the following manner. The 
propagator is
\bea\label{twopointproton}
S^{\alpha\beta}(p,d)=i\int dx e^{ipx}
\la 0|TP^\alpha(x) 
\bar{P}^\beta(0)|0\ra\co
\eea
where $P(x)$ denotes the proton field. The $Z$-factor is obtained from
\bea\label{eqzfactorproton}
\left\{\lim_{p^2\rightarrow m_p^2}
(m_p-\not\! p)S(p,d)\right\}u(p,r)
=Z_N u(p,r)\scs d>4\co\nn
\eea
where $m_p$ denotes the physical proton mass, $u(p,r)$ is a spinor 
in $d$ dimensions with $(\not\!\! p-m_p)u(p,r)=0$, and where $Z_N$ depends 
on the prescription used for performing the integration  over 
the Feynman parameters (standard, regular or infrared). We obtain for the 
infrared regularized $Z$-factor at order $p^2$ 
\bea\label{Z-N}
Z_N&=&1+
 \frac{e^2}{8\pi^2}\,\,(C_{\rm IR} - C_{\rm UV}) 
\nonumber\\[2mm]
&-&\frac{g_A^2 m^2}{64\,\pi^2\,F^2}\,
\left(H^I_0+\frac{\Delta_\pi}{m^2}\,H^I_\pi\right)+O(p^3)\,,
\eea
with
\bea
H_0^I  &=&
r^2\,(6+9\,C_{\rm UV}+18\,\ln r) \co\nn
H_\pi^I  &=&
 \,-(5 +3\,C_{\rm UV}  +6\,\ln{r}) \sem \quad
r=M_\pi/m\fs\nn
\eea
Here, we have replaced the physical nucleon mass with its chiral limit mass 
$m$ - this is correct at the order of the low--energy expansion we are working. 
We will do this in all loop amplitudes considered below. 
\begin{figure}[thb]
\begin{center}
\resizebox{0.4\textwidth}{!}{\includegraphics*{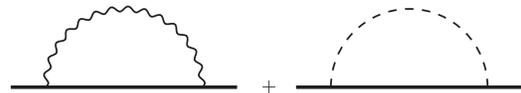}}
\end{center}

\vskip-1.cm

\caption{Photon and pion loop contribution to the  two--point function
(\protect{\ref{twopointproton}}) of the proton. Neutral and 
charged pions are running in the loop.}
\label{fig:nucleonGF}
\end{figure}

\setcounter{equation}{0}
\section{Evaluation of $S$-matrix elements}
\label{sec:smatrix}
Here, we  describe the  evaluation of the $S$-matrix element for the process 
$\pi^- p\rightarrow \pi^- p$ in this framework. The basic element to be 
evaluated first is the truncated four-point function
\bea
\langle 0|T\pi^-(x_1)\pi^+(x_2)P^\alpha(x_3)\bar{P}^\beta(x_4)|0\rangle
\eea
in $d$ space-time dimensions. We take into account tree graphs evaluated with  
${\cal L}_{eff}$ and one--loop graphs generated by ${\cal L}_\pi^{(p^2)}$, 
${\cal L}_\pi^{(e^2)}$ and ${\cal L}_N^{(p)}$. The relevant diagrams are 
displayed in Figs. (\ref{fig:Born},\ref{fig:counterterms},\ref{fig:vector},
\ref{fig:axial},\ref{fig:strong}). At the end, one multiplies the result with 
the incoming and outgoing spinors for the proton field, and incorporates the 
contribution from the wave function renormalization constants for the pion 
and nucleon field, determined in sections \ref{sec:twopoint_pi},
\ref{sec:twopoint_N} in the standard manner. For general momenta of the incoming 
and outgoing particles, the limit $d\rightarrow 4$ does not exist due to the 
infrared divergences which are generated by virtual photons. As we discussed in 
section~\ref{sec:kinematics}, these divergences are absent in the amplitude at 
threshold, provided that one has removed the infrared divergent Coulomb phase. 
Here, we are interested in the infrared finite quantity ${\cal T}_{\pi N}$ 
defined in Eq.~(\ref{thr}). The final result depends on the manner the 
integration over the Feynman parameters is carried out (standard, regular or 
infrared). Here, we work with the prescription given in Ref.~\cite{BL1}, which 
preserves chiral power counting. In order to keep track of UV and IR divergences,
we found it convenient to evaluate the regular part of each  diagram separately, 
and subtract it from the same diagrams calculated in the standard manner. The 
procedure was illustrated  in sections~\ref{sec:twopoint_pi},
\ref{sec:twopoint_N} for the two--point functions. It can  straightforwardly be 
generalized to any of the one--loop diagrams encountered here. The schematic 
prescription is as follows:

\begin{itemize}

\item[i)]
Graphs with no virtual nucleons need no infrared regularization - graphs with 
closed nucleon lines vanish in infrared regularization.
 
\item[ii)]
Graphs with pions (or photons or both) and nucleons running in the loop, are
evaluated in the following manner. Heavy (nucleons)  and light (pions and 
photons) lines in the diagrams are combined together separately by using 
Feynman parameterization,  as  described in section 6 of Ref.~\cite{BL1}.
The pertinent denominators are then combined with a single Feynman parameter 
[$x$ attached to the heavy and $(1-x)$ to the light line]. The infrared 
regularization consists~\cite{BL1} in extending the region of integration 
over this parameter, from $[0,1]$ to $[0,\infty]$.

\item[iii)]
\begin{sloppypar}
In order to distinguish between ultraviolet and infrared divergences in graphs 
with a virtual photon, we evaluate the infrared regularized part as the 
difference between the standard and regular contribution. Symbolically, 
\end{sloppypar}
\bea\label{iruv}
M^I&=&\int_0^\infty dx F=M^S-M^R\co\nn
M^S&=&\int_0^1 dx F\scs \quad
 M^R=-\int_1^\infty dx F\fs
\eea
 Here, $M^I$ stands for the final result of the calculation of any diagram in 
the infrared regularization, and $F$ denotes the corresponding integrand, where 
the integration over the remaining Feynman parameters is implicit.

\item[iv)]
One may clearly distinguish UV and IR divergences in the one--loop diagrams
which are contained in $M^S$, see sections~\ref{sec:twopoint_pi},
\ref{sec:twopoint_N}. All divergences in $M^R$ at $d\to4$ are declared to be 
ultraviolet. At the end, all infrared divergences cancel in ${\cal T}_{\pi N}$. 
Ultraviolet divergences of the one--loop diagrams cancel against the divergent 
parts of the low--energy constants, as a result of which the quantity 
${\cal T}_{\pi N}$ is IR and UV finite. 

\end{itemize}  
The calculations may be simplified considerably if one makes use of the fact
that power counting is now preserved: Nucleon (meson) propagators count as 
$p^{-1}$ ($p^{-2}$). The numerators can then be simplified by dropping terms 
that generate contributions beyond the order $p^3$ considered here. The 
procedure is described in~\cite{BL2}. Here, one may use in addition that we 
work at threshold, where the pion momenta are proportional to the nucleon 
momenta, $p'^\mu=p^\mu,q'^\mu=q^\mu=p^\mu M_{\pi}/{m_p}$. Examples of such
simplifications are given in section~\ref{sec:triangle}, where the
triangle diagram is evaluated.

We add a comment on the quark mass difference $\Delta=m_d-m_u$ that ~occurs ~in 
~the ~isospin ~violating part of the threshold amplitude. It shows up in two 
ways: first, through the neutron and proton masses in Feynman diagrams, and 
second, 
through  vertexes in the effective Lagrangian. Consider first the latter 
contributions. From  Eq. ~(\ref{Born}), it is explicitly seen that $\Delta$ does 
not occur at order $p^3$, once the physical masses are used in the tree diagram 
Fig.\ref{fig:Born}b. Further, in Eq.~(\ref{Born}), we may replace the neutron
mass by the proton mass at order $p^3$ - $\Delta$ still does not occur. Next 
consider loop diagrams, which are evaluated with the nucleon masses in the 
chiral limit, $m_p=m_n=m$, and are of order $p^3$ in the framework used here. 
Therefore, $\Delta$ cannot contribute. [Note that there is a difference between 
the nucleon and pion masses in this respect: The electromagnetic contribution 
to the charged pion mass is of the same order in the chiral counting as the 
leading term. Therefore, it matters what pion  is running in the loop.] We 
conclude that the amplitude ${\cal T}^m$ in (\ref{sym}) is of order 
$M_{\pi}^2$ and therefore beyond the accuracy of the calculation considered here.

\setcounter{equation}{0}
\section{The triangle diagram}
\label{sec:triangle}

In ~this ~section ~we ~investigate ~the ~infrared-divergent Coulomb phase, as 
well as the singular terms in the real part of the amplitude that behave like
$|{\bf p}|^{-1}$ in the vicinity of the physical threshold. At one loop,
these factors arise when one calculates radiative corrections to the strong
tree--level amplitudes which are obtained from the nucleon Lagrangian 
${\cal L}_N^{(p)}$. For illustration, we consider the strong diagram emerging 
from the Weinberg-Tomozawa (WT) vertex (Fig.~\ref{fig:Born}a). The scattering 
amplitude corresponding to this diagram is
\eq\label{WT}
T_{WT}=\bar u(p')\, \frac{\not\! q+\not\! q'}{4F^2}\, u(p)\, .
\en
The Coulomb singularities arise when one considers the virtual photon corrections
to this vertex with a particular topology $(v_1)$, depicted in 
Fig.~\ref{fig:vector}. There are two diagrams of this type, with the photon
emitted and absorbed either in the initial or in the final state. The
contributions of these diagrams to the amplitude at threshold are the same,
so it suffices to consider one of them, and multiply the result by $2$.
The total contribution of these diagrams is
\eq\label{V1}
&&T_{v_1}=\frac{e^2}{2F^2}\int\frac{d^dk}{(2\pi)^di}\,
\frac{\bar u(p')(\not\! q+\not\! q'-\not\! k)}{[-k^2]}
\nonumber\\[2mm]
&&
\times
\frac{(m+\not\! p+\not\! k)
(2\not\! q-\not\! k)u(p)}
{[m^2-(p+k)^2]\, [M_{\pi}^2-(q-k)^2]}\, .
\en
The numerator can be rewritten in the following manner,
\eq\label{num_3}
&&\bar u(p')(\not\! q+\not\! q'-\not\! k)
(m+\not\! p+\not\! k)
(2\not\! q-\not\! k)u(p) =
\nonumber\\[2mm]
&&\bar u(p')\biggl\{ (2(\not\! q+\not\! q')-\not\! k)\, [m^2-(p+k)^2]
\nonumber\\[2mm]
&&
+4\not\! q'(m+\not\! p)\not\! q+
4\not\! q'\not\! k\not\! q\biggr\}u(p)\, .
\en
According to the discussion in section~\ref{sec:smatrix}, we drop here the last 
term that does not contribute to the amplitude at $O(e^2p)$. Then, after some 
algebra, one arrives at the following expression for the sum of contributions 
of tree (WT) and one--loop $(v_1)$ diagrams to the invariant amplitude $D$ 
given by Eq.~(\ref{AB}),
\eq\label{D_vec}
D_{WT}&+&D_{v_1}=\frac{s-u}{8mF^2}\,\biggl\{1+e^2
\biggr[\frac{J_1}{2M_{\pi}^2}
\nonumber\\[2mm]
&+&4J_2+4(s-m^2)J_\gamma(s)\biggr]
\biggr\}+\cdots\, , 
\en
where the ellipsis stands for higher-order terms in the chiral expansion.
The scalar integrals that enter the expression (\ref{D_vec}), are given by
\eq\label{J}
J_1&=&\int\frac{d^dk}{(2\pi)^di}\,\frac{1}{M_{\pi}^2-k^2}
=\frac{M_{\pi}^2}{16\pi^2}\, C_{\rm UV}^\pi\, ,
\nonumber\\[2mm]
J_2&=&\int\frac{d^dk}{(2\pi)^di}\,\frac{1}{[-k^2]\,[M_{\pi}^2-(q-k)^2]}
\nonumber\\[2mm]
&=&-\frac{1}{16\pi^2}\,(C_{\rm UV}^\pi-1)\, ,
\nonumber\\[2mm]
J_\gamma(s)&=&\int\frac{d^dk}{(2\pi)^di}\,\frac{1}
{[-k^2][m^2-(p+k)^2]\,[M_{\pi}^2-(q-k)^2]}\, .\nn
\en
The representation (\ref{D_vec}) turns out to be very convenient to study the
threshold behavior of the amplitude. The reason for this is that the 
integral $J_\gamma(s)$ enters here with a coefficient which counts at
$O(e^2p^2)$ - so, the regular part of $J_\gamma(s)$, which is a polynomial in
the external momenta and masses, cannot contribute at $O(e^2p)$ to the
amplitude $D$. Consequently, one may use the standard dimensional
regularization instead of infrared regularization for calculating this
integral. 

Introducing Feynman parameters, the momentum integration in $J_\gamma(s)$ can
be explicitly done,
\eq\label{Jgamma}
J_\gamma(s)&=&\frac{\Gamma(3-\frac{d}{2})}{(4\pi)^{d/2}}\,
\int_0^1d\alpha_1d\alpha_2d\alpha_3\,
\nonumber\\[2mm]
&\times&
\delta\biggl(1-\sum_{i=1}^3\alpha_i\biggr)\, [J(\alpha)]^{d/2-3}\, ,
\nonumber\\[2mm]
J(\alpha)&=&(\alpha_1+\alpha_2)(\alpha_1m^2+\alpha_2M_{\pi}^2)
-\alpha_1\alpha_2s\, .
\en
One may single out the infrared divergence at $d=4$ by rescaling the Feynman 
parameters according to $\alpha_{1,2}\to(1-\alpha_3)\alpha_{1,2}$ - the 
integration over the parameter $\alpha_3$ then factorizes, and we arrive at the 
following expression,
\eq\label{Jgamma_1}
&&J_\gamma(s)=\frac{1}{32\pi^2m^2}\,
\biggl\{ (C_{\rm IR}+1)\int_0^1\frac{d\alpha_1}{R}
+\int_0^1\frac{d\alpha_1\ln R}{R}\biggr\}\, ,
\nonumber\\[2mm]
&&R=\alpha_1+(1-\alpha_1)r^2
-\alpha_1(1-\alpha_1)\bar s\, ,\quad
\bar s=\frac{s}{m^2}\, .
\en
Using the formulae (\ref{R-int_2}), (\ref{R-int_4}) from 
appendix~\ref{app:basic}, and expanding near threshold, for the sum 
$D_{WT}+D_{v_1}$ we finally obtain
\eq\label{WTphase}
D_{WT}&+&D_{v1}=\frac{M_{\pi}}{2F^2}\,
\biggl\{1+
2i\alpha\theta_C(|{\bf p}|)
+\frac{\pi\alpha\mu_c}{|{\bf p}|}
\nonumber\\
&+&\frac{\alpha}{\pi r}\, 
I_V^1+O(|{\bf p}|)
\biggr\}\, ,
\nonumber\\[2mm]
I_V^1&=&-\frac{r}{8}\,(7C_{\rm UV}+8C_{\rm IR}
+30\ln r+16)+O(r^2)
\sem 
\nn
 r&=&\frac{M_\pi}{m}\co
\en
where the Coulomb phase $\theta_C(|{\bf p}|)$ is given in Eq.~(\ref{theta_C}). 

From Eq.~(\ref{WTphase}) it is evident, that the combination 
$\exp^{-2i\alpha\theta_C(|{\bf p}|)}\,(D_{WT}+D_{v1})$ [cf with
Eq.~(\ref{thr})] does not contain the Coulomb phase at $O(\alpha)$. Further,
dropping the term proportional to $|{\bf p}|^{-1}$, one may read off 
the corresponding contribution $I_V^1$ to the threshold amplitude 
${\cal T}_{\pi N}$, see table~\ref{tab:V}, entry $(v_1)$. Finally, we note that 
in the amplitude there is no term that behaves like $\ln|{\bf p}|$ near
threshold - the coefficient $B_2$ in Eq.~(\ref{thr}) first appears at two--loop 
order~\cite{Bern4}.

\setcounter{equation}{0}
\section{The threshold amplitude at order $p^3$}
\label{sec:results}

We do not provide any further details of the calculations that can be done along 
the lines illustrated in previous sections. In this section, we combine the 
various pieces and present the final expression for the threshold scattering 
amplitude ${\cal T}_{\pi N}$. This quantity is given by a sum of several 
contributions: to the contribution of Born diagrams Eq.~(\ref{Born}), 
multiplied by the pion and nucleon wave function renormalization factors 
(Eqs.~(\ref{Z-pi}) and (\ref{Z-N}), respectively), one has to add the 
contributions from vector (Fig.~\ref{fig:vector}), axial (Fig.~\ref{fig:axial}) 
and strong (Fig.~\ref{fig:strong}) loops, as well as the counterterm contribution
${\cal T}^{\rm ct}$ displayed in~(\ref{c.t.}),
\eq\label{TpiN_ini}
{\cal T}_{\pi N}=Z_\pi Z_N\, {\cal T}^{\rm B}+{\cal T}^{\rm V}
+{\cal T}^{\rm A}+{\cal T}^{\rm S}
+{\cal T}^{\rm ct}\, ,
\en
where
\eq\label{sums}
{\cal T}^{\rm V}&=&\frac{e^2 m}{8\pi^2 F^2}\,
\sum_{i=1}^6I^i_V\, ,
\qquad
{\cal T}^{\rm A}=\frac{e^2 g_A^2m}{32\pi^2 F^2}\,
\sum_{i=1}^9I^i_A\, ,
\nonumber\\[2mm]
{\cal T}^{\rm S}&=&\frac{m^3}{\pi^2F^4}\,
\sum_{i=1}^{24}(I_0^i+\frac{\Delta_\pi}{m^2}\, I_\pi^i)\, .
\en
The quantities $I_V^i$, $I_A^i$, $I_0^i$ and $I_\pi^i$, that correspond to the 
contributions from individual vector, axial and strong diagrams, are listed in
tables~\ref{tab:V}, \ref{tab:A} and \ref{tab:I0_Ipi} in
appendix~\ref{app:piN-tables}. Not all of 
the diagrams that are shown in Figs.~\ref{fig:vector}, \ref{fig:axial} and 
\ref{fig:strong}, do contribute to the threshold amplitude at ~$O(p^3)$. The 
diagram $(s_{24})$ in Fig.~\ref{fig:strong} e.g. vanishes trivially after the 
momentum integration. Moreover, there are two reasons for which some of the 
diagrams that are formally $O(p^3)$ in chiral counting, start to contribute at
higher order: 
\begin{itemize}
\item[i)]
In the one-particle reducible diagrams Fig.~\ref{fig:axial}: $(a_1)$, $(a_2)$, 
$(a_5)$, and Fig.~\ref{fig:strong}:
$(s_1)$, $(s_2)$, $(s_5)$, $(s_6)$, $(s_7)$, $(s_8)$, $(s_9)$, $(s_{17})$,
~the ~neutron ~propagator is followed by $\not\! q_\pi\gamma_5 u(p)$ (initial
state), or preceded by $\bar u(p')\not\! q_\pi\gamma_5$ (final state),
where $q_\pi$ stands either for $q$ or for $q'$. Formally, the 
combination [$\not\! q_\pi\gamma_5 \times$ propagator] is of chiral 
order zero. However, putting $\gamma_5$ through the nucleon propagator changes 
the sign of the nucleon momentum, and the whole expression starts to contribute 
at $O(p)$.
\item[ii)]
Doing the simplification of numerators in a manner described in
sections~\ref{sec:smatrix} and \ref{sec:triangle} for the diagrams
$(a_3)$, $(a_6)$, $(a_7)$ and $(s_3)$, $(s_4)$, $(s_{10})$, $(s_{11})$, it is 
easy to observe that the leading-order contributions in the numerators vanish 
at threshold.
\end{itemize}

For this reason, in tables~\ref{tab:V}, \ref{tab:A} and \ref{tab:I0_Ipi} we 
do not display the contributions from the above-mentioned diagrams.

Adding all pieces together in the amplitude, we have checked that all UV 
divergences, as expected, cancel with the divergent parts of the LECs, whereas 
the IR divergences cancel among themselves. The final results is thus given in 
terms of renormalized LECs $l_i^r$, $k_i^r$, $d_i^r$, $g_i^r$. Further, we split 
the threshold amplitude in its isospin-conserving and isospin-violating parts
according to Eq.~(\ref{spl}). The expressions for these parts are given by (for 
convenience, we use here the physical proton mass $m_p$ and the physical pion 
decay constant $F_\pi$)
\eq\label{IC}
&&{\cal T}_{\pi N}^0=
\frac{M_\pi}{2F_\pi^2}-\frac{g_A^2M_\pi^2}{4m_pF_\pi^2}
+\frac{2M_\pi^2}{F_\pi^2}\,(-2c_1+c_2+c_3)
\nonumber\\[2mm]
&&+\frac{g_A^2M_\pi^3}{8m_p^2F_\pi^2}
+\,\frac{M_\pi^3}{16\pi^2F_\pi^4}\,\biggl(1+\frac{3\pi g_A^2}{4}
-2\,\ln\frac{M_\pi}{\mu}\biggr)
\nonumber\\[2mm]
&&+\frac{M_\pi^3}{F_\pi^4}\, l_4^r +
\frac{4M_\pi^3}{F_\pi^2}\,(d_1^r+d_2^r+d_3^r+2d_5^r)
+ O(p^4)\, ,
\\[5mm]
\label{IV}
&&\delta{\cal T}=\delta{\cal T}_2+\delta{\cal T}_3+O(p^4)\, ,
\nonumber\\[2mm]
&&\delta{\cal T}_3=\delta{\cal T}_3^{\rm str}+\delta{\cal T}_3^{\rm em}
+\delta{\cal T}_3^{\rm ct}\, ,
\\[2mm]
&&\delta{\cal T}_2=\frac{4\Delta_\pi}{F_\pi^2}\, c_1
-\frac{e^2}{2}\,(4f_1+f_2)\, ,
\nonumber\\[2mm]
&&\delta{\cal T}_3^{\rm str}=
-\frac{M_\pi\Delta_\pi}{32\pi^2F_\pi^4}\,
\biggl(3+
\frac{33\pi g_A^2}{4}+2\,\ln\frac{M_\pi}{\mu}\biggr)\, ,
\nonumber\\[2mm]
&&\delta{\cal T}_3^{\rm em}=
-\frac{e^2M_\pi g_A^2}{32\pi^2F_\pi^2}\,
\biggl(2+\pi+8\,\ln2+12\,\ln\frac{M_\pi}{\mu}\biggr)\, ,
\nonumber\\[2mm]
\label{IVct}
&&\delta{\cal T}_3^{\rm ct}=
-\frac{8M_\pi\Delta_\pi}{F_\pi^2}\, d_5^r
\nonumber\\[2mm]
&&
+2e^2M_\pi\biggl(g_6^r+g_8^r
-\frac{5}{9F_\pi^2}\,(k_1^r+k_2^r)\biggr)\, ,
\en
where $\delta{\cal T}_2$, $\delta{\cal T}_3$ are of order $p^2$ and $p^3$,
respectively. From the above equations one may easily check that the amplitude 
does not depend on the scale $\mu$. Note that here we have replaced the pion 
decay constant in the chiral limit $F$ by the physical decay constant $F_\pi$,
\eq\label{F_pi}
\hspace*{-5.mm}
F_\pi=F\biggl\{1+\frac{M_\pi^2}{F^2}\,\biggl(l_4^r
-\frac{1}{8\pi^2}\ln\frac{M_\pi}{\mu}\biggr)
+O(M_\pi^4)\biggr\}\, .
\en
This is the reason that $l_4^r$ occurs in Eq.~(\ref{IC}).

Equations (\ref{IV}) and (\ref{IVct}) represent our main result. This is a 
complete calculation of the isospin--breaking part of the threshold scattering 
amplitude at $O(p^3)$ in relativistic baryon ChPT, by using the infrared 
regularization method of Refs.~\cite{BL1,BL2}. The isospin-conserving part of 
the amplitude, given in Eq.~(\ref{IC}), agrees with the one displayed in 
appendix~\ref{app:lagrtables} of Ref.~\cite{Fettes:2000xg}. A calculation of isospin breaking 
effects at order $p^3$ has been performed  by the same authors in~\cite{Fettes}.
Since they do not provide a  complete analytic expression of the scattering 
amplitude, it is not possible to compare the results.

\setcounter{equation}{0}
\section{The size of the low--energy constants}
\label{sec:LECs}

In this section, we  discuss the size of the low--energy constants that occur 
in the isospin breaking amplitude $\delta{\cal T}$. The amplitude 
$\delta {{\cal T}_2} \, \, (\delta {{\cal T}}_3)$ contains 
 $4f_1+f_2$, $c_1$ ($k_1^r+k_2^r$, $g_6^r+g_8^r$, $d_5^r$). We start the 
discussion with  $\delta{\cal T}_2$ and consider the constant $c_1$. It has been 
determined from threshold data~\cite{Koch} in Ref.~\cite{BL2}. The uncertainty 
in $c_1$ may be obtained from equation (20.2) in~\cite{BL2}, using H\"ohler's  
values~\cite{Hoehler} for the uncertainties in the threshold parameters and in 
the coupling constant $g_{\pi N}$. We find\footnote{We use for 
$M_\pi,m_p,\ldots$ in the following the values quoted in~\cite{Groom:in}, 
in particular $F_\pi=92.4$ MeV, $|g_A|$=1.267.}
\bea\label{c1value}
c_1 =-(0.93\pm 0.07 ){\rm GeV}^{-1}\, .
\eea
For a comparison with other determinations of $c_1$, see table 1 
in Ref.~\cite{Buttiker}.

Next, consider $f_1$ and $f_2$ that occur in the order $e^2$ pion-nucleon 
Lagrangian. Up to terms of order ${\mathbold{\pi}}^4$, that Lagrangian is
\bea\label{f1f2}
{\cal{L}}_N^{(e^2)}&=&F^2\bar{\Psi}O\Psi\co\nonumber\\
O&=&\frac{e^2}{2}[(f_1+f_3)\cdot 1_2+f_2\tau_3]
\nonumber\\[2mm]
&-&\frac{2f_1e^2}{F^2}\,\pi^+\pi^-\cdot 1_2+ef_2\bar{Q}+\cdots\co
\nonumber\\
\bar{Q}&=&\frac{e}{4F^2}\,\left(\begin{array}{ll}
             -2\pi^+\pi^-&\sqrt{2}\pi^0\pi^+\\
             \sqrt{2}\pi^0\pi^-&2\pi^+\pi^-\end{array}
     \right);\,\, 
1_2={\rm diag}(1,1)\, ,\nn
\eea
where $\tau_3$ denotes the Pauli-matrix. From this decomposition, it is seen 
that $f_1$ occurs in the  chiral expansion of the nucleon mass and  in elastic 
pion-nucleon scattering $\pi^\pm p(n)\rightarrow \pi^\pm p(n)$. The 
electromagnetic part of the proton-neutron mass difference is given by the 
constant $f_2$ at leading order in the chiral expansion,
\bea
-e^2F^2f_2=(m_p-m_n)^{\rm em}\fs
\eea
Here, we disagree with the result Eq. (12) of Ref.~\cite{Steininger} by a 
factor of 2. Numerically, we use $(m_p-m_n)^{\rm em}=(0.76\pm0.3)$ 
MeV~\cite{physrep}, or
\bea\label{f2protonneutron}
f_2&=&-(0.97\pm 0.38){\mbox{GeV}}^{-1}\fs
\eea
We are now left with the determination of $f_1$. The sum $m_p+m_n$ contains 
the combination $e^2(f_1+f_3)$ -  the constants $f_1$ and $f_3$  can 
therefore not be disentangled from information on the nucleon masses. 
We may consider $m_p+m_n$  as a quantity that fixes $f_3$, once $f_1$ is known. 
Therefore, elastic pion-nucleon scattering is the only realistically 
accessible source of information on $f_1$. In principle, one may consider 
combinations of amplitudes that vanish in the isospin symmetry limit, and 
determine $f_1$ from those. The combination 
\bea\label{f1determination}
X=T^{\pi^+p\rightarrow \pi^+p}+T^{\pi^-p\rightarrow\pi^-p}
-2T^{\pi^0p\rightarrow\pi^0p}
\eea
has this property. The tree graphs of $X$ start at order $p^2$ and contain 
$f_1$ - that one may try to determine hence from here. Of course, one is 
faced with a problem of accuracy: in order to determine $X$, one needs to 
consider the difference of two large numbers, quite aside from the fact that 
the cross section $\pi^0p\rightarrow\pi^0p$ is not known experimentally. It 
remains to be seen whether a combination of experimental data and lattice 
calculations could resolve the problem also in practice.

In the absence of precise experimental information  on $f_1$, we can i) rely 
on order-of-magnitude estimates, or ii) consider model calculations. As to 
order-of-magnitude estimates, we follow Fettes and 
Mei\ss ner~\cite{Fettes:2000vm} and write
\bea\label{f1ordermeissner}
F^2e^2|f_1|\simeq \frac{\alpha}{2\pi}m_p\co\nonumber
\eea
or
\bea\label{f1estimate}
|f_1|\simeq 1.4{\rm {GeV}}^{-1}\co
\eea
 because $f_1$ is due to a genuine  photon loop at the quark level (we divide 
by $2\pi$ rather than by 4$\pi$~\cite{Fettes:2000vm} to be on the conservative 
side). This estimate also confirms the expectation~\cite{Bern3}  that $|f_1|$ 
has the same size as $|f_2|$, see Eq.~(\ref{f2protonneutron}). As to model 
calculations, we refer the reader to Ref.~\cite{Faessler}, where $c_1$ and 
$f_{1,2,3}$ have been determined in a quark model, with the result
\bea\label{c1faessler}
c_1&=&-1.2{\rm GeV}^{-1}\scs 
\nonumber\\[2mm]
F^2(f_1,f_2,f_3)&=&(-19.5\pm 1.6,
-8.7\pm 0.7,18\pm 1.5)\, {\rm MeV}\fs\nn
\eea
Finally, we come to the low--energy constants that occur in the 
next--to--leading order amplitude $\delta{\cal T}_3$. The contributions of the 
relevant LECs to $\delta_\epsilon$ is suppressed by one power of the pion mass 
with respect to the ones from $c_1, f_1$ and $f_2$ that occur at leading 
order - therefore, we expect their effect to be substantially smaller, because 
they represent polynomial parts of the  amplitudes, not chiral singular pieces. 
 The  value for $d_5^r$ is given in table 1 of Ref.~\cite{BL2} (we have chosen 
the entry with the largest uncertainty assigned)
\bea\label{d5values}
16M_\pi^2d_5^r(\mu)=0.04\pm 0.06 
-\frac{2M_\pi^2}
{F^2}\,l_4^r(\mu)\scs ~~\mu=1{\rm GeV}\fs\nn
\eea
Here, we have translated the $d_5^r$ from Ref.~\cite{BL2} into the present 
scheme according to  Eq.~(\ref{d5new}). Below, we will use
 $l_4^r(1{\mbox{GeV}})=2.9\cdot 10^{-3}$~\cite{CGL}. For $k_1^r+k_2^r$, we 
invoke standard dimensional arguments for an estimate of its size, 
$|k_i^r|\simeq \frac{1}{16\pi^2}$, and apply  the same rule to $g_6^r+g_8^r$,
\bea\label{k12g68}
|k_1^r+k_2^r|\simeq \frac{2}{16\pi^2}
\scs
|g_6^r+g_8^r|\simeq\frac{2}{16\pi^2 F^2}\fs
\eea
[In Ref.~\cite{Fettes},  $g_6^r+g_8^r$ was estimated from an analysis of 
pion-nucleon scattering at low energies. However, in that reference, 
renormalization was treated in a manner different from the framework used here, 
and it is not clear to us how we can compare the couplings determined there 
with the ones used here, see appendix \ref{app:Berezinian}.] As a check of 
this procedure, we apply the same estimate to $d_5^r$, namely 
$|d_5^r| \simeq \frac{1}{16\pi^2F^2}$. Comparing with Eq. (\ref{d5values}),
we see that this guess is a rather generous one.

\setcounter{equation}{0}
\section{Ground-state energy-level shift}
\label{sec:energyshift}

We are now ready to provide numerical values for  the correction 
$\delta_\epsilon$ in the strong energy-level shift $\delta_\epsilon$
according to Eqs.~(\ref{LO}) and (\ref{deltaE}). In addition to the LECs, 
we need the value for the scattering lengths $a_{0^+}^\pm$. We take from
Ref.~\cite{PSI}
\bea
a_{0^+}^+ +a_{0^+}^-=0.0883M_\pi^{-1}\fs
\eea
\subsection{The correction $\delta_\epsilon$}
The contribution from the leading term $\delta{\cal T}_2$ was already determined
in Ref.~\cite{Bern3}. This contribution contains the three LECs $f_1$, $f_2$ and
$c_1$. For $f_1$,  the estimate $|f_1|<|f_2|$ was used in~\cite{Bern3}. The 
corresponding contribution to $\delta_\epsilon$ is $\pm 1.9\cdot 10^{-2}$. Here, 
we use the estimate (\ref{f1estimate}), which is slightly more conservative,
and display in table \ref{tab:numerics} the contribution due to 
$\delta{\cal T}_2|_{f_1=0}, f_1$ and to the bound state correction $K$. We will 
use below this information to estimate the uncertainty in the final result.

\begin{table}[t]
\caption{
Individual contributions to $\delta_\epsilon$. $K$ denotes the
bound-state correction in Eq.~(\protect{\ref{deltaE}}). 
See text for details.}\label{tab:depsilon}
\label{tab:numerics}
\def\arraystretch{2.0}
\vspace*{.2cm}
\begin{center}
\begin{tabular}{|c||c|c|}
\hline
source &\multicolumn{2}{c|}{$\delta_\epsilon\times 10^2$}\\
\hline
\hline
$\delta{\cal T}_2|_{f_1=0}$
& \multicolumn{2}{|c|} {$-5.5$} \\
$f_1$&\multicolumn{2}{|c|}{$\pm 2.8$}\\
\hline
    & $\mu=500$ MeV & $\mu=1~{\rm GeV}$ \\
\hline
$\delta{\cal T}_3^{\rm str}$& $-3.5$ & $-3.4$ \\
\hline
$\delta{\cal T}_3^{\rm em}$ & $0.4$ & $1.1$ \\
\hline
$d_5^r$&$-0.3\pm 0.3$&$-0.2\pm 0.3$\\
\hline
$k_1^r+k_2^r$&\multicolumn{2}{c|}{$\pm 0.2$}\\[-2mm]
$g_6^r+g_8^r$&\multicolumn{2}{c|}{$\pm 0.4$}\\
\hline
 $K$&\multicolumn{2}{|c|}{$0.66$}\\
\hline
 vac. pol.~\cite{Eiras}& \multicolumn{2}{|c|} {$0.48$} \\
\hline
\end{tabular}
\end{center}
\end{table}

The next order in the chiral expansion of $\delta_\epsilon$ is obtained 
from the second term in (\ref{deltaE}), evaluated  with 
$\delta{\cal T}= \delta{\cal T}_3$. We split this term further into 
contributions that stem from the photon loop, strong loops and counterterms, 
as indicated in (\ref{IV}). Note that the individual terms are scale dependent,
whereas the full result is not, of course. In the table, we present the values 
obtained from the strong loops and photon loops at two values of the scale, 
$\mu=500$ MeV and $\mu$ = 1 GeV. The contributions from the strong loops are 
substantial and of negative sign. This large contribution is mainly due to the 
three graphs Fig.~\ref{fig:strong} $(s_{19}), (s_{21})$ and $(s_{22})$,  
which add
\bea\label{triangle}
\delta{\cal T}_3^{\rm str}\biggr|_{(s_{19}),(s_{21}),(s_{22})} 
= -\frac{33\pi g_A^2}{128\pi^2 
F_\pi^4}M_\pi\Delta_\pi\co
\eea
or $\delta_\epsilon=-3.5\cdot 10^{-2}$. We find it amusing to see that it is 
the triangle graph that generates these large contributions. [In the self-energy 
type graph $(s_{19})$, one has to expand the neutral pion mass in the 
propagator around the charged pion mass. This expansion generates a 
triangle-type graph.] A graph with triangle-topology was found to generate a 
large contribution to the photo production of neutral pions a long time 
ago~\cite{photo}. In addition, that contribution is responsible for the
breakdown of a so-called low--energy theorem in that process. Further, this 
graph also plays an important role in the low--energy analysis of pion-nucleon 
scattering~\cite{BL1,BL2}.

One might worry that the next--to--leading order correction could be large as 
compared to the leading order, and jeopardize the chiral expansion also in 
the isospin breaking sector. However, as we already noticed, the reason for 
this large correction is well understood - it is due to the triangle graphs. 
We do therefore not consider this to be a problem.

Finally, we display in the table  the contributions from the counterterms in 
$\delta{\cal T}^{\rm ct}$, using the estimates displayed above. It is seen that 
the effect of the LECs at this order is indeed suppressed with respect to the 
leading order ones. We then add the central values at the scale $\mu=M_\rho$, 
with $f_1=g_i^r=k_i^r=0$. For the uncertainty, we add in quadrature the 
contributions from $f_1$, 
$\delta f_2,\delta c_1$, $k_1^r+k_2^r\,,\, g_6^r+g_8^r$ and from $\delta d_5^r$.
In this manner, we obtain
\bea\label{final}
\delta_\epsilon=(-7.2\pm 2.9)\cdot 10^{-2} \fs
\eea
This is our final result for the correction $\delta_\epsilon$ . The uncertainty 
in (\ref{final}) is dominated by the largely unknown coupling $f_1$, as is seen 
from the third row in table \ref{tab:numerics}. It does not take into account 
contributions from order $p^4$ in the chiral expansion of the isospin breaking 
amplitude $\delta{\cal T}$.

We note that our numerical results for the isospin--breaking corrections in the 
amplitude cannot be directly confronted with those of Ref.~\cite{Fettes}, which 
refer to different physical quantities. Numerical values for the 
isospin--breaking effects given in Ref.~\cite{Fettes} are typically of order 
$\sim 1-2~\%$.

\subsection{Comparison with model calculations}

Here we compare our result with the potential model calculation performed in 
Ref.~\cite{Sigg}, and with the evaluation~\cite{Faessler} of the couplings
at leading order in a quark model, Eq. ~(\ref{c1faessler}). We start with
the latter. The central  values in (\ref{c1faessler}) lead\footnote{
In Ref.~\cite{Faessler} slightly different values for the pion masses were
used. The result quoted there for the correction to the energy shift
corresponds to those values. We thank V. Lyubovitskij for correspondence.} -  
adding the bound state correction $K$ - to $\delta_\epsilon=-2\cdot10^{-2}$ 
~at ~order ~$p^2$. ~Taken ~at ~face ~value, this determines the coupling 
$f_1$ in our calculation: $f_1=-1.43{\rm GeV}^{-1}$, in agreement with the 
estimate (\ref{f1estimate}). This leads finally to 
$\delta_\epsilon=-4.3\cdot 10^{-2}$ at order $p^3$.

\begin{figure}[H]
\begin{center}
\resizebox{0.4\textwidth}{!}{\includegraphics*{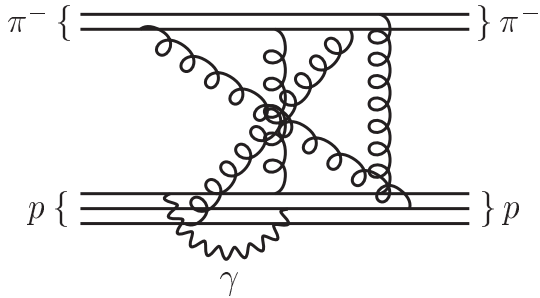}}
\end{center}
\caption{An example of a QCD diagram, which is not taken into account in
the potential model of Ref.~\cite{Sigg}. Wiggly lines denote gluons.}
\label{fig:QCD}
\end{figure}

We now come to the potential model discussed ~in ~Ref.~\cite{Sigg}, which
 predicts
\bea\label{potential}
\delta_\epsilon=(-2.1\pm0.5)\cdot10^{-2}\fs
\eea
We have the following comments.
\begin{itemize}
\item[-]
The result~(\ref{potential}) looks considerably more precise than the 
effective field theory evaluation. There are no unknown LECs occurring in the 
framework used in Ref. \cite{Sigg}. 
\item[-]
The leading terms in the effective field theory framework - that lead to the 
large uncertainty in (\ref{final}) - are on the other hand  not all 
incorporated~\cite{Bern3} in (\ref{potential}). An example of a QCD diagram, 
whose contribution is completely omitted, is shown in Fig.~\ref{fig:QCD}. 
In the potential model~\cite{Sigg}, it can emerge neither from the 
$n\gamma$ channel (there is no intermediate state with a neutron in
Fig.~\ref{fig:QCD}), nor from the Coulomb rescattering (the photon in
Fig.~\ref{fig:QCD} is attached only to the proton). At the level of the
effective Lagrangians, such contributions are encoded in the low--energy
constants $f_1$ and $f_2$.
\item[-]
The potential model used in Ref.~\cite{Sigg} takes into account 
terms at order $p^3,$ $p^4,\cdots$ in the language of the low--energy 
expansion. We are not aware of a proof that it includes all of them.
On the other hand, one can construct potentials that certainly 
do~\cite{potential-PLB}. Unfortunately, these potentials do not determine 
the LECs, because the LECs are used instead to pin down the potentials.
\item[-]
The observation that potential models do not, in general, include all effects
of QCD+QED is not new. Aside from the hadronic atom
case~\cite{old-pipi,Bern1,Bern2,Bern4}, we mention here the investigation of
Fettes and Mei\ss ner in $\pi N$ scattering~\cite{Fettes}. These authors 
pointed out that graphs like the one displayed in 
Fig.~\ref{fig:vector}~$(v_6)$ generate a very large effect, that has not been 
fully accounted for in existing phase shift analyses.
\item[-]
We conclude that the uncertainty in (\ref{potential}) is underestimated - it 
does not reflect the systematic errors inherent in the method.
\end{itemize}

\setcounter{equation}{0}
\section{Summary and conclusions}
\label{sec:concl}

\begin{enumerate}
\item
The ~aim ~of ~the ~present work was to evaluate the ground--state energy of  
pionic hydrogen in the framework of QCD+QED. We have performed the calculation 
at next--to--leading order in the low--energy expansion, relying on the method 
of effective field theories. According to the investigations of 
Ref.~\cite{Bern3}, all that is needed for this purpose is an evaluation of the 
elastic $\pi N$ scattering amplitude at order $p^3$.
\item
In order to achieve this goal, we have invoked relativistic baryon ChPT in 
manifestly Lorentz invariant form~\cite{BL1}, and generalized the procedure  
to allow for virtual photons. To have a powerful check on our calculation, we 
have in addition performed a heat-kernel evaluation of the ultraviolet 
divergences at order $p^3$, using a consistent set of low--energy meson--meson 
and meson--nucleon effective Lagrangians that are used in the calculation of the 
$\pi N$ scattering amplitudes at $O(p^3)$.
\item
We have then calculated the $\pi^-p$ elastic scattering amplitude at 
threshold - in the presence of isospin breaking - at $O(p^3)$ in the 
low--energy expansion. The result contains unitarity corrections and 
counterterm contributions generated by LECs. The quark mass difference 
$m_d-m_u$ only enters at order $p^4$, which is not considered here.
\item
At leading order, the contribution to the energy shift is generated~\cite{Bern3} 
by three counterterms $f_1,f_2$ and $c_1$. Whereas the latter two can be 
determined from other sources, the size of $f_1$ needs yet to be determined in a
model-independent manner.
\item
The loop contributions, which contain mass splitting effects from diagrams with 
strong loops, turn out to be sizeable. Graphs with a particular topology 
(triangle graphs) turn out to be particularly  important  - their contribution 
to the energy shift indeed is large and negative, a sizeable fraction of the 
leading order term. Graphs with the same topology play an important role e.g. 
in photo production of neutral pions~\cite{photo}.
\item
The LECs at next--to--leading order are suppressed by one power of $M_{\pi}$ 
and thus expected to have a small effect on the  energy shift. This
expectation turns out to be correct for one of the couplings that has
recently been determined~\cite{BL2} in a comprehensive analysis of $\pi N$
scattering. Estimating the size of the remaining terms with dimensional
arguments, it indeed turns out that their contribution is about an order of
magnitude smaller than the leading order result. The final result is given in 
Eq.~(\ref{final}) 
\item
Whereas it is true that we cannot yet provide a reliable error estimate of the 
final result, because $f_1$ is not known with sufficient accuracy, it is also 
true that the calculation is systematic: The result (\ref{IC})-(\ref{IVct}) for 
the threshold amplitude allows one to evaluate the energy-level shift (at this 
order in the low--energy expansion) from first principles, once one has worked 
out a reliable estimate for the LECs.
\item
A precise determination of the scattering lengths $a_{0^+}^+ + a_{0^+}^-$ 
from a precise measurement of the ground-state energy-level shift of pionic 
hydrogen has to await a more precise determination of $f_1$ in our opinion.
This fact is hidden in the potential model calculation~\cite{Sigg}, 
that quotes a very small uncertainty. As we outlined above, this result
does not reflect all systematic uncertainties hidden in this approach:
potentials in general do not incorporate the constraints from QCD+QED, unless 
one imposes these constraints on them~\cite{potential-PLB}. A method different 
from effective field theories to perform this matching is not in sight. We 
conclude that one is bound to know the LECs, quite independently of the 
framework used.
\end{enumerate}

\begin{acknowledgement}
We thank T.~Becher, G.~Colangelo, G.~Ecker, T.~Ericson, 
B. Kubis, H.~Leutwyler, V.E. Lyubovitskij, 
H. Neufeld, U.-G. Me\-i\ss ner, and M.E. Sainio
 for interesting discussions, and G. Ecker, U-G. Mei\ss ner and H. Neufeld
for useful comments on the manuscript.
The work was fulfilled while M.A.I., E.L. and M.M.
visited the University of Bern. 
This work was supported in part by the Swiss National Science
Foundation, and by TMR, ~BBW-Contract ~No. 97.0131  ~and  ~EC-Contract
~No. ~ERBFMRX - CT980169 (EU\-RO\-DA\-$\Phi$NE), and by SCOPES Project
No. 7UZPJ65677. 
M.A.I. appreciates the partial support by the Russian Fund
of Basic Research under Grant No. 01-02-17200.
\end{acknowledgement}

\appendix

\renewcommand{\thesection}{\Alph{section}}
\renewcommand{\theequation}{\Alph{section}\arabic{equation}}

\setcounter{equation}{0}
\section{Meson--meson and meson--nucleon Lagrangians}
\label{app:lagrtables}
In this appendix, we give the operator basis and the divergent parts of the 
LECs of the meson--meson and meson--nucleon effective chiral Lagrangians, that 
were used in the calculations. We set the charge matrices to their physical, 
constant values,

\bea
{\cal Q}=\frac{e}{3}\,{\rm diag}\,(2,-1)\scs \qquad
Q=e\,{\rm diag}\,(1,0)\, ,
\eea
and use the following notation,
\eq\label{not}
&&[D_\mu,X]\doteq \partial_\mu X+[\Gamma_\mu,X]\, ,
\nonumber\\[2mm]
&&\bar\Psi\, (O+{\rm h.c.})\,\Psi\doteq\bar\Psi\, O\,\Psi+{\rm h.c.}\, ,
\nonumber\\[2mm]
&&\bar\Psi\, i(O-{\rm h.c.})\,\Psi\doteq i(\bar\Psi\, O\,\Psi-{\rm h.c.})\, ,
\en
where $X$ denotes any bosonic operator, and $\Psi$ is a fer\-mi\-o\-nic 
field. We use the notation of Bjorken-Drell~\cite{Bjorken} for the Dirac
matrices.

\newpage

\begin{table}[H]
\caption{
Operator basis and the divergent parts of the low--energy couplings 
in the $O(p^4)$ meson Lagrangian~\cite{GL} and
in the $O(e^2p^2)$ meson Lagrangian~\cite{KU}. The terms that do not 
contain pion fields, and terms of order $e^4$ are not displayed.}
\label{tab:lagr_p4}

\begin{center}
\def\arraystretch{1.4}
\begin{tabular}{|r|c|c|}
\hline\hline
$i$ & $O_i^{(p^4)}$ & $\gamma_i$ \\
\hline
$1$&$\frac{1}{4}\,\langle d^\mu U^\dagger d_\mu U\rangle^2$ & $\frac{1}{3}$\\
$2$&$\frac{1}{4}\,\langle d^\mu U^\dagger d^\nu U\rangle\langle d_\mu
U^\dagger d_\nu U\rangle$ & $\frac{2}{3}$\\
$3$&$\frac{1}{16}\,\langle\chi^\dagger U+U^\dagger\chi\rangle^2$ & $-\frac{1}{2}$\\
$4$&$\frac{1}{4}\,\langle d^\mu U^\dagger d_\mu\chi+d^\mu\chi^\dagger d_\mu
U\rangle$ & $2$\\ 
$5$&$\langle {\cal R}^{\mu\nu}U{\cal L}_{\mu\nu}U^\dagger\rangle$ & 
$-\frac{1}{6}$\\
$6$&$\frac{i}{2}\,\langle {\cal R}^{\mu\nu}d_\mu Ud_\nu U^\dagger$ 
& $-\frac{1}{3}$\\
& $+{\cal L}^{\mu\nu}d_\mu U^\dagger d_\nu U\rangle$ & \\
$7$&$-\frac{1}{16}\,\langle\chi^\dagger U-U^\dagger\chi\rangle^2$ & $0$\\

\hline\hline
$i$ & $O_i^{(e^2p^2)}$ & $\sigma_i$ \\
\hline
$1$&$\langle d^\mu U^\dagger d_\mu U\rangle\langle{\cal Q}^2\rangle$ & $-\frac{27}{20}-\frac{1}{5}\,Z$\\
$2$&$\langle d^\mu U^\dagger d_\mu U\rangle
\langle{\cal Q}U{\cal Q}U^\dagger\rangle$ & $2Z$\\
$3$&$\langle d^\mu U^\dagger{\cal Q}U\rangle
\langle d_\mu U^\dagger{\cal Q}U\rangle$ & $-\frac{3}{4}$
\\
&$+\langle d^\mu U{\cal Q}U^\dagger\rangle
\langle d_\mu U{\cal Q}U^\dagger\rangle$ & \\
$4$&$\langle d^\mu U^\dagger{\cal Q}U\rangle
\langle d_\mu U{\cal Q}U^\dagger\rangle$ & $2Z$\\
$5$&$\langle\chi^\dagger U+U^\dagger\chi\rangle\langle{\cal Q}^2\rangle$ & $-\frac{1}{4}-\frac{1}{5}\,Z$\\
$6$&$\langle\chi^\dagger U+U^\dagger\chi\rangle
\langle{\cal Q}U{\cal Q}U^\dagger\rangle$ & $\frac{1}{4}+2Z$ \\
$7$&$\langle(\chi U^\dagger+U\chi^\dagger){\cal Q}$
& $0$ \\
&$+(\chi^\dagger U+U^\dagger\chi){\cal Q}\rangle\langle{\cal Q}\rangle$ & \\
$8$&$\langle(\chi U^\dagger-U\chi^\dagger){\cal Q}U{\cal Q}U^\dagger$
&$\frac{1}{8}-Z$ \\
&$+(\chi^\dagger U-U^\dagger\chi){\cal Q}U^\dagger{\cal Q}U\rangle$ & 
\\
$9$&$\langle d_\mu U^\dagger[{\cal C}_R^\mu,{\cal Q}]U+
d_\mu U[{\cal C}_L^\mu,{\cal Q}]U^\dagger\rangle$ & $\frac{1}{4}$\\
$10$&$\langle {\cal C}_R^\mu U{\cal C}_{L\mu}U^\dagger\rangle$ & $0$\\
\hline
\end{tabular}
\end{center}

\end{table}

\begin{table}[H]
\caption{
Operator basis in the $O(p^2)$ pion-nucleon Lagrangian~\cite{I}, and
in the $O(e^2)$ pion-nucleon Lagrangian~\cite{Muller}.}
\label{tab:LN_p2}

\begin{center}
\def\arraystretch{1.4}
\begin{tabular}{|r|c|c|}
\hline\hline
$i$ & $O_{i}^{(p^2)}$ & $O_{i}^{(e^2)}$\\
\hline
$1$&$\langle\chi_+\rangle$ & $\langle\hat Q_+^2-Q_-^2\rangle$ \\
$2$&$-\frac{1}{4m^2}\,\langle u_\mu u_\nu\rangle
(D^\mu D^\nu+\mbox{h.c.})$ & $\langle Q_+\rangle\hat Q_+$ \\
$3$&$\frac{1}{2}\,\langle u_\mu u^\mu\rangle$ 
& $\langle\hat Q_+^2+Q_-^2\rangle$ \\
$4$&$\frac{i}{4}\,\sigma^{\mu\nu}[u_\mu,u_\nu]$&\\
$5$&$\hat\chi_+$&\\
$6$&$\frac{1}{8m}\,\sigma^{\mu\nu}F^+_{\mu\nu}$&\\
$7$&$\frac{1}{8m}\,\sigma^{\mu\nu}\langle F^+_{\mu\nu}\rangle$&\\

\hline 

\end{tabular}
\end{center}

\end{table}

\newpage

\begin{table}[H]
\caption{
Operator basis and the divergent parts of the low--energy couplings in
the third-order strong pion-nucleon Lagrangian. The 
operator basis is the same as
the one used in Ref.~\cite{I} for the relativistic operators. The 
divergent parts
are different from Ref.~\cite{I}, since we use the EOM to eliminate additional
operators that arise in the HBChPT formulation, and work with a different
basis in the mesonic sector. For comparison with Ref.~\cite{EM}, see 
appendix~\ref{app:Berezinian}.}
\label{tab:LN_p3}
\begin{center}
\def\arraystretch{1.4}
\begin{tabular}{|r|c|c|}
\hline\hline
$i$ & $O_i^{(p^3)}$ & $\beta_i$ \\
\hline
$1$ & $-\frac{1}{2m}\,([u_\mu,[D_\nu,u^\mu]]D^\nu+\mbox{h.c.})$ &
$-\frac{1}{6}\, g_A^4$ \\
$2$ & $-\frac{1}{2m}\,([u_\mu,[D^\mu,u_\nu]]D^\nu+\mbox{h.c.})$ &
$-\frac{1}{12}-\frac{5}{12}\, g_A^2$ \\ 
$3$ & $\frac{1}{12m^3}\,([u_\mu,[D_\nu,u_\lambda]]$ &
$\frac{1}{2}+\frac{1}{6}\, g_A^4$ \\
&$\times(D^\mu D^\nu D^\lambda
+\mbox{sym})+\mbox{h.c.})$
& \\ 
$4$ & $-\frac{1}{2m}\,(\varepsilon^{\mu\nu\alpha\beta}\langle
u_\mu u_\nu u_\alpha\rangle D_\beta+\mbox{h.c.})$& $0$ \\
$5$ & $\frac{i}{2m}\,([\chi_-,u_\mu]D^\mu-\mbox{h.c.})$ &
$-\frac{5}{24}+\frac{5}{24}\, g_A^2$ \\
$6$ & $\frac{i}{2m}\,([D^\mu,\hat F^+_{\mu\nu}]D^\nu-\mbox{h.c.})$ &
$-\frac{1}{6}-\frac{5}{6}\, g_A^2$ \\
$7$ & $\frac{i}{2m}\,([D^\mu,\langle F^+_{\mu\nu}\rangle]D^\nu-\mbox{h.c.})$
& $0$ \\
$8$ & $\frac{i}{2m}\,(\varepsilon^{\mu\nu\alpha\beta}\langle
\hat F^+_{\mu\nu}u_\alpha\rangle D_\beta-\mbox{h.c.})$ & $0$ \\
$9$ & $\frac{i}{2m}\,(\varepsilon^{\mu\nu\alpha\beta}\langle
F^+_{\mu\nu}\rangle u_\alpha D_\beta-\mbox{h.c.})$ & $0$ \\
$10$ & $\frac{1}{2}\,\gamma^\mu\gamma_5u_\mu\langle u^\nu u_\nu\rangle$ &
$\frac{1}{2}\, g_A+\frac{5}{2}\, g_A^3$\\
& & $+2g_A^5$ \\
$11$ & $\frac{1}{2}\,\gamma^\mu\gamma_5u^\nu\langle u_\mu u_\nu\rangle$ & 
$\frac{1}{2}\, g_A-\frac{3}{2}\, g_A^3$\\
& & $-\frac{2}{3}\, g_A^5$ \\
$12$ & $-\frac{1}{8m^2}\,(\gamma^\mu\gamma_5u_\mu
\langle u_\nu u_\lambda\rangle$ & $-2g_A-g_A^3$ \\
& $\times\{D^\nu,D^\lambda\}+\mbox{h.c.})$ & 
 $-2g_A^5$ \\
$13$ & $-\frac{1}{8m^2}\,(\gamma^\mu\gamma_5u_\lambda
\langle u_\mu u_\nu\rangle$ & $g_A^3+\frac{2}{3}\, g_A^5$   \\
& $\times \{D^\nu,D^\lambda\}+\mbox{h.c.})$ & 
\\
$14$ & $\frac{i}{4m}\,(\sigma^{\mu\nu}\langle u_\nu[D_\lambda,u_\mu]\rangle
D^\lambda-\mbox{h.c.})$ & 
$\frac{1}{3} g_A^4$ \\
$15$ & $\frac{i}{4m}\,(\sigma^{\mu\nu}\langle u_\mu[D_\nu,u_\lambda]\rangle
D^\lambda-\mbox{h.c.})$ & 
$0$ \\
$16$ & $\frac{1}{2}\,\gamma^\mu\gamma_5\langle\chi_+\rangle u_\mu$ & 
$\frac{1}{2}\, g_A+g_A^3$ \\
$17$ & $\frac{1}{2}\,\gamma^\mu\gamma_5\langle\chi_+ u_\mu\rangle$ & 
$0$ \\
$18$ & $\frac{i}{2}\,\gamma^\mu\gamma_5[D_\mu,\chi_-]$ & $g_A$ \\
$19$ & $\frac{i}{2}\,\gamma^\mu\gamma_5[D_\mu,\langle\chi_-\rangle]$ &
$-\frac{1}{2}\, g_A$ \\
$20$ & $-\frac{i}{8m^2}\,(\gamma^\mu\gamma_5[\hat F^+_{\mu\nu},u_\lambda]$ & 
$g_A+g_A^3$ \\
& $\times\{ D^\nu,D^\lambda\}-\mbox{h.c.})$ &  \\
$21$ & $\frac{i}{2}\,\gamma^\mu\gamma_5
[\hat F^+_{\mu\nu},u^\nu]$ & $-g_A^3$ \\
$22$ & $\frac{1}{2}\,\gamma^\mu\gamma_5[D^\nu,F^-_{\mu\nu}]$ & $0$ \\
$23$ & $\frac{1}{2}\,\varepsilon^{\mu\nu\alpha\beta}\gamma_\mu\gamma_5
\langle u_\nu F^-_{\alpha\beta}\rangle$ & $0$ \\
\hline
\end{tabular}
\end{center}

\end{table}

\newpage

\begin{table}[H]
\caption{
Operator basis and the divergent parts of the low--energy couplings in
the third-order electromagnetic pion-nucleon Lagrangian. The operator basis is 
the same as the one used in Ref.~\cite{thesis} for the relativistic
operators (modulo signs and factors of $i$), 
whereas the divergent parts are different from those
displayed in Ref.~\cite{thesis}, see appendix \ref{app:Berezinian}.}

\label{tab:LN_e2p}

\begin{center}
\def\arraystretch{1.4}
\begin{tabular}{|r|c|c|}
\hline\hline
$i$ & $O_i^{(e^2p)}$ & $\eta_i$ \\
\hline
$1$ & $\frac{1}{2}\langle Q_+^2-Q_-^2\rangle\gamma^\mu\gamma_5 u_\mu$ &
$g_A(2+g_A^2 $\\
& & $+4Z+12g_A^2Z)$\\
$2$ & $\frac{1}{2}\langle Q_+\rangle^2\gamma^\mu\gamma_5 u_\mu$ &
$-\frac{1}{2}\, g_A(3+2g_A^2$\\
& & $+4Z+12g_A^2Z)$ \\ 
$3$ & $\frac{1}{2}\,\gamma^\mu\gamma_5\langle Q_+\rangle
\langle Q_+u_\mu\rangle$ &
$\frac{1}{2}\, g_A(1-4Z$\\
& & $+4g_A^2Z)$ \\
$4$ & $\frac{1}{2}\,\gamma^\mu\gamma_5 Q_+\langle Q_+u_\mu\rangle$ &
$g_A(-1+4Z $\\
& & $-4g_A^2Z)$ \\
$5$ & $\frac{1}{2}\,\gamma^\mu\gamma_5 Q_-\langle Q_-u_\mu\rangle$ &
$g_A(3-2g_A^2$ \\
& & $-4Z+4g_A^2Z)$\\
$6$ & $\frac{i}{2m}\, \langle Q_+\rangle\langle Q_-u_\mu\rangle 
D^\mu+\mbox{h.c.}$
& $ \frac{3}{4}-3g_A^2$\\
& & $ -\frac{1}{3}\, Z-\frac{5}{3}\, g_A^2Z$ \\
$7$ & $\frac{i}{2m}\, Q_-\langle Q_+u_\mu\rangle 
D^\mu+\mbox{h.c.}$
& $-\frac{9}{2}-\frac{2}{3}\, Z$\\
& & $-\frac{10}{3}\, g_A^2Z$ \\
$8$ & $\frac{i}{2m}\, Q_+\langle Q_-u_\mu\rangle 
D^\mu+\mbox{h.c.}$
& $-\frac{3}{2}+6g_A^2$ \\
& & $+\frac{2}{3}\, Z+\frac{10}{3}\, g_A^2Z$ \\
$9$ & $-\frac{1}{2m}\,[Q_+,c^+_\mu]D^\mu+\mbox{h.c.}$ & $-2$ \\
$10$ & $-\frac{1}{2m}\,[Q_-,c^-_\mu]D^\mu+\mbox{h.c.}$ & 
$-\frac{1}{2}+\frac{9}{2}\, g_A^2$ \\
$11$ & $\frac{i}{2}\gamma^\mu\gamma_5[Q_+,c^-_\mu]$ & $-g_A$ \\
$12$ & $\frac{i}{2}\gamma^\mu\gamma_5[Q_-,c^+_\mu]$ & $g_A$ \\
\hline
\end{tabular}
\end{center}

\end{table}

\setcounter{equation}{0}
\section{Divergent counterterms}
\label{app:Berezinian}

It is common and useful in ChPT to identify the divergent parts of
counterterms not just within a calculation of a particular process, but rather
in full generality, using the background field and heat-kernel methods.
One of the benefits of such an approach is that it provides a nontrivial check
for any particular calculation - the loop infinities must be cancelled by
the a priori known counterterms.

In spite of the fact that the method of Ball~\cite{Ball} can accommodate large
variety of regularizations, including the dimensional one, it is not
straightforward to extend it to cover also the Becher-Leutwyler infrared
regularization~\cite{BL1,BL2}. One may, however, utilize the fact that
divergences encountered in the infrared regularization are the same as 
those occurring in heavy baryon ChPT ~\cite{BL1,BL2}.

In HBChPT the baryon field is split into velocity-dependent ''heavy'' and
''light'' components\footnote{We follow the tradition of denoting the 
velocity by the same symbol $v_\mu$ which was used for the external vector
field. The meaning should be always clear from the context.}
\eq \label{HB}
N_{v}\left( x\right) &=& e^{imv\cdot x}P_{v}^{+}\Psi\left(  x\right)
\, , \nonumber \\[2mm]
H_{v}\left( x\right) &=& e^{imv\cdot x}P_{v}^{-}\Psi\left(  x\right) 
\, , \nonumber \\[2mm]
P_{v}^{\pm} &=&\frac{1}{2}\left(  1\pm\gamma^{\mu}v_{\mu}\right)\quad\, ;   
\qquad
v^{2}=1\, ,
\en
and the heavy component $H_{v}\left( x\right)$ is integrated out. The
propagator of the remaining light component is a homogeneous function of the
residual momentum $k_{\mu}=p_{\mu}-mv_{\mu}$, leading to a consistent chiral
power counting in dimensionally regularized HBChPT.

The heat-kernel method in HBChPT leads to a lengthy and ~cumbersome 
~calculation. ~The ~amount of work required can be reduced significantly by 
introduction of the so-called super-heat-kernel method in recent work of Neufeld 
{\it et al}~\cite{Berezinian,Neufeld}. Following closely the procedure of 
Neufeld, we worked out the complete divergence structure of the one--loop 
generating functional corresponding to the Lagrangian
\eq \label{L_HB}
{\cal L}={\cal L}_{HB}^{(p)}+{\cal L}_{\pi}^{(p^{2})}+{\cal L}
_{\pi}^{(e^{2})} +{\cal L}_\gamma\, ,
\en
where
\eq \label{L_HB_1}
{\cal L}_{HB}^{(p)}=\overline{N}_{v}\left(  iv^{\mu}D_{\mu}+g_{A}S^{\mu
}u_{\mu}\right)  N_{v}
\en
is the HB projection of ${\cal L}_{N}^{(p)}$, with $S^{\mu}=\frac{i}{2}
\gamma_{5}\sigma^{\mu\nu}v_{\nu}$.

The starting point of the background field machinery is a decomposition of
fields in the Lagrangian into background fields and fluctuations.
Here one has to distinguish between two different elements of the chiral
coset space ~$SU\left( 2\right) _{L}\times SU\left( 2\right) _{R}/$ 
$SU\left(2\right) _{V}$, namely $u_{L}$ and $u_{R}$, occurring in the  
definitions of various fields appearing in chiral invariant Lagrangians,
\eq \label{LR_fields}
\hspace*{-5.mm}U&=&u_{R}u_{L}^{\dagger}
\, ,
\nonumber\\[2mm]
\hspace*{-5.mm}
u_{\mu}&=&i\left[  u_{R}^{\dagger}\left(  \partial_{\mu}-iR_{\mu}\right)
u_{R}
-u_{L}^{\dagger}\left(  \partial_{\mu}-iL_{\mu}\right)  u_{L}\right]
\, , 
\nonumber\\[2mm]
\hspace*{-5.mm}
\Gamma_{\mu}&=&\frac{1}{2}\left[ u_{R}^{\dagger}\left( \partial_{\mu}
-iR_{\mu}\right) u_{R}
+u_{L}^{\dagger}\left( \partial_{\mu}-iL_{\mu}
\right)  u_{L}\right]\, .
\en
Decomposing the fields into classical background fields and fluctuations
\eq \label{fluct}
u_{R}&=&u_{\mathrm{cl}}e^{i\mathbold{\xi}
/2F}\qquad\, ,\qquad
u_{L}=u_{\mathrm{cl}}^{\dagger}
e^{-i\mathbold{\xi}/2F}\, ,
\nonumber \\[2mm]
N&=&N_{\mathrm{cl}}+\eta \qquad\, ,\qquad
A^{\mu}=A_{\mathrm{cl}}^{\mu}
+\epsilon^{\mu} \, ,
\en
and switching to the Euclidean formulation of the theory, which is more
convenient when dealing with photons, one first merges the bosonic fields 
into 7- component objects
$\tilde{\zeta}$, 
\eq \label{xi_tilde}
\hspace*{-5.mm}
\tilde{\zeta}^{T}=\left(  \xi^{1},\xi^{2},\xi^{3},\epsilon^{0},\epsilon
^{1},\epsilon^{2},\epsilon^{3}\right) 
\scs
{\mathbold{\xi}}=\sum \xi^i{\mathbold{\tau}^i}\fs
\en
At the next step, one expands the action up to second order in the fluctuations 
$\tilde\zeta$ and $\eta$ (Euclidean analog of the formula (3.1) in 
Ref.~\cite{Neufeld}). Once the explicit form of the fluctuations is worked out,
one can use (the Euclidean version of) Neufeld's master equations (3.19)-(3.21), 
Ref.~\cite{Neufeld}. At the end, the results are  transformed back to Minkowski 
space.

After some straightforward, yet tedious, algebra one finds the divergent
counterterm Lagrangian in a ''raw'' form,
\eq \label{raw}
{\cal L}^{\rm{raw}}&=&{\cal L}_{\pi}^{\rm{raw}}
+{\cal L}_{\pi\gamma}^{\rm{raw}}
+{\cal L}_{N}^{\rm{raw}}+{\cal L}_{N\gamma}^{\rm{raw}}\, ,
\en
\bea \label{L_pi_raw}
\cal{L}_{\pi}^{\rm{raw}} &=&\frac{1}{16\pi^{2}\left( d-4\right)}
\left\{  \frac{1}{12}\left\langle u\cdot u\right\rangle ^{2}
+\frac{1}{6}
\left\langle u_{\mu}u_{\nu}\right\rangle \left\langle u^{\mu}u^{\nu}
\right\rangle \right.
\nonumber\\[2mm]
&+&\left.\frac{3}{32}\left\langle \chi_{+}\right\rangle ^{2}\right.
-\frac{1}{12}\left\langle \hat {\cal F}_{+\mu\nu} \hat {\cal F}_{+}^{\mu\nu}
\right\rangle 
\nonumber\\[2mm]
&-&\frac{i}{12}\left\langle \left[ u_{\mu},u_{\nu}\right]
\hat {\cal F}_{+}^{\mu\nu}\right\rangle 
+\frac{1}{4}\left\langle u\cdot u\right\rangle \left\langle
\chi_{+}\right\rangle \biggr\} \, ,
\eea
\bea \label{L_pigamma_raw}
\cal{L}_{\pi\gamma}^{\rm{raw}}&=&\frac{F^{2}}{16\pi^{2}\left(
d-4\right)  }
\left\{  -\frac{27+4Z}{20}\left\langle u\cdot u\right\rangle
\left\langle {\cal Q}^{2}\right\rangle \right.
\nonumber\\[2mm]
&+&\left.2Z\left\langle u\cdot
u\right\rangle \left\langle {\cal Q}_{+}^{2}-{\cal Q}_{-}^{2}
\right\rangle \right. 
\nonumber \\
&+&\left.\frac{3}{2}\left(  \left\langle u_{\mu}{\cal Q}_{+}\right\rangle
^{2}+\left\langle u_{\mu}{\cal Q}_{-}\right\rangle ^{2}\right)\right.
\nonumber\\[2mm]
&+& \left. 2Z\left(
\left\langle u_{\mu}{\cal Q}_{+}\right\rangle ^{2}-\left\langle u_{\mu
}{\cal Q}_{-}\right\rangle ^{2}\right)\right. 
\nonumber \\
&-&\left. \frac{5+4Z}{20}\left\langle \chi_{+}\right\rangle \left\langle
{\cal Q}^{2}\right\rangle \right.
+\frac{1+8Z}{4}\left\langle \chi_{+}\right\rangle
\left\langle {\cal Q}_{+}^{2}-{\cal Q}_{-}^{2}\right\rangle 
\nonumber \\
&-&\left. \frac{i}{2}\left\langle u_{\mu}\left[ \cal{C}_{-}^{\mu},{\cal Q}
_{+}\right]  +u_{\mu}\left[ \cal{C}_{+}^{\mu},{\cal Q}_{-}\right]
\right\rangle\right.
\nonumber\\[2mm]
&+&\left.\frac{75-80Z+168Z^{2}}{50}F^{2}\left\langle {\cal Q}
^{2}\right\rangle ^{2}\right.
\nonumber \\
&-&\left. \frac{15+2Z+12Z^{2}}{5}F^{2}
\left\langle{\cal Q}_{+}^{2}-
{\cal Q}_{-}^{2}\right\rangle \left\langle {\cal Q}^{2}\right\rangle\right. 
\nonumber \\
&+&\left. \frac{3+4Z+24Z^{2}}{2}F^{2}\left\langle {\cal Q}_{+}^{2}-
{\cal Q}_{-}^{2}\right\rangle ^{2}\right.
\nonumber\\[2mm]
&+&\left.\frac{i}{2}
\left\langle \left[  D_{\mu},u^{\mu}\right]  
\left[ {\cal Q}_{-},{\cal Q}_{+}\right]  \right\rangle \biggr\}\, ,\right.
\eea
\bea\label{raw-N}
{\cal L}_{N}^{{\rm raw}} 
&=&\frac{1}{16\pi^{2}\left(  d-4\right)}\frac{1}{F^2}\bar{N}
\biggl\{ \left.\frac{g_A^4}{8}\left[ u_{\mu},\left[iv\cdot D ,u^{\mu}\right] 
\right]\right.
\nonumber\\[2mm]
&-&\left.\frac{1+5g_A^2}{12}\left[u_{\mu},\left[i D ^{\mu},v\cdot u\right]
\right]
\right. 
\nonumber  \\[2mm]
&+&
\left.\frac{4-g_A^4}{8}\left[ v\cdot u,\left[ iv\cdot D ,v\cdot u\right] 
\right]\right.
\nonumber\\[2mm] 
&+&
\left. \frac{4g_A-g_A^5}{8}S\cdot u\left\langle u\cdot u\right\rangle
\right.
\nonumber \\[2mm]
&+&\left.\frac{6g_A-6g_A^3+g_A^5}{12}u_{\mu}\left\langle u^{\mu} 
S\cdot u\right\rangle \right.
\nonumber\\[2mm]
&+&\left.\frac{-8g_A+g_A^5}{8}S\cdot u\left\langle \left(v\cdot u\right) ^{2}
\right\rangle  \right.
\nonumber\\
&-&\left.\frac{g_A^5}{12}v\cdot u\left\langle v\cdot uS\cdot u\right\rangle 
\right.
\left.+\frac{4g_A-g_A^3}{8}S\cdot u\left\langle \chi _{+}\right\rangle 
\right.
\nonumber\\[2mm]
&-&\left.\frac{1+5g_A^2}{6}\left[ D _{\mu},\hat F_{+}^{\mu \nu}\right] v_{\nu}
\right.
\left.+g_A i S_{\mu}v_{\nu}\left[ \hat F_{+}^{\mu \nu},v\cdot u\right] 
\right.
\nonumber\\[2mm]
&-&\left.\frac{4g_A^3+3g_A^5}{16}i v_{\lambda}
\varepsilon ^{\lambda \mu \nu \rho}
\left\langle u_{\mu}u_{\nu}u_{\rho}\right\rangle\right.  
\nonumber\\[2mm]
&-&\left.\frac{g_A^4}{4}[S^\mu,S^\nu]\left\langle u_{\mu}
\left[iv\cdot  D,u_{\nu }\right] \right\rangle \right.
\nonumber\\[2mm]
&-&\left.\frac{1+5g_A^2}{12}\left[\left[i D_{\mu},u^{\mu}\right],v\cdot 
u\right]
\right.
\nonumber\\[2mm]
&-&\left.\frac{g_A^3}{4}v_{\lambda }\varepsilon ^{\lambda \mu \nu \rho}
\left\langle \hat F_{+\mu \nu }u_{\rho }\right\rangle
\right.
\left.-3g_A^2i\left(v\cdot  D \right) ^3\right.
\nonumber\\[2mm]
&+&
\left.g_A^3 v\cdot \overleftarrow{D} S\cdot u v\cdot \overrightarrow{D}
\right.
\nonumber\\[2mm]
&-&\left.\frac{12g_A^2+9g_A^4}{16}\right.
\left.\left( \left\langle u\cdot u\right\rangle
i v\cdot  D +\mathrm{h.c.}\right) 
\right.
\nonumber\\[2mm]
&+&\left.\frac{8+9g_A^4}{16}\right.
\left.\left( \left\langle
\left( v\cdot u\right) ^{2}\right\rangle iv\cdot  D +\mathrm{h.c.}\right)
\right.
\nonumber\\
&+&\left.\frac{g_A^3}{3}\left( \left[ v\cdot  D ,S\cdot u\right] v\cdot  D 
+{\rm h.c.}\right) \right.
\nonumber\\[2mm]
&-&\left.\frac{9g_A^2}{16}\left( \left\langle \chi_{+}
\right\rangle i v\cdot  D +{\rm h.c.}\right)\right.
\nonumber\\[2mm]
&+&\left. \frac{4g_A^2+g_A^4}{4}([S^\mu,S^\nu]u_{\mu}u_{\nu}
i v\cdot  D +{\rm h.c.})\right. 
\nonumber\\[2mm]
&+&\left. g_A^2( [S^\mu,S^\nu]\hat F_{+\mu \nu}
v\cdot  D 
+{\rm h.c.}) \biggr\} N\, ,\right.
\eea
\bea\label{raw-Ngamma}
{\cal L}_{N\gamma}^{\rm{raw}}&=&\frac{1}{16\pi^{2}\left(  d-4\right)}
\bar{N}
\left\{  \frac{g_{A}}{2}\left(  8Z-g_{A}^{2}\right)  \left\langle
Q_{+}^{2}-Q_{-}^{2}\right\rangle S\cdot u\right. 
\nonumber\\
&-& \frac{g_{A}}{2}\left(  4+4Z-g_{A}^{2}\right)  
\left\langle Q_{+}\right\rangle ^{2}S\cdot u
\nonumber\\[2mm]
&-&2g_{A}
\left( 1+Z-Zg_{A}^{2}\right)  \left\langle
Q_{+}S\cdot u\right\rangle \left\langle Q_{+}\right\rangle 
\nonumber\\[2mm]
&+& 4Zg_{A}\left(  1-g_{A}^{2}\right)  \left\langle Q_{+}S\cdot u\right\rangle
Q_{+}
\nonumber\\[2mm]
&+&2g_{A}\left(  1-g_{A}^{2}\right)  \left(  1-2Z\right)  \left\langle
Q_{-}S\cdot u\right\rangle Q_{-}
\nonumber\\[2mm]
&+& \left(  1-3g_{A}^{2}\right)  \left\langle Q_{-}v\cdot u\right\rangle
\left\langle Q_{+}\right\rangle 
-4\left\langle Q_{+}v\cdot u\right\rangle
Q_{-}
\nonumber\\[2mm]
&-&2\left(  1-3g_{A}^{2}\right)  \left\langle Q_{-}v\cdot u\right\rangle
Q_{+}
- 2\left[  Q_{+},iv\cdot c_{+}\right] 
\nonumber\\[2mm]
&-&\frac{1}{2}\left(  1-9g_{A}^{2}\right)
\left[  Q_{-},iv\cdot c_{-}\right] 
\nonumber\\[2mm]
&-&2\left(  \left\langle Q_{+}\right\rangle
Q_{+}iv\cdot D+\mathrm{h.c.}\right) 
\nonumber\\
&-& \frac{1}{4}\left(  2+3g_{A}^{2}+24Zg_{A}^{2}\right) 
 \left(  \left\langle
Q_{+}^{2}-Q_{-}^{2}\right\rangle iv\cdot D+\mathrm{h.c.}\right) 
\nonumber\\[2mm]
&+& \frac{1}{4}\left( 2 + 3 g_{A}^{2} + 12 Zg_{A}^{2}\right)
\left.(  \left\langle Q_{+}\right\rangle ^{2}iv\cdot D
+\mathrm{h.c.})
\right\}  N \, . \nn
\eea

The results for $\cal{L}_{\pi}^{\rm{raw}}$ and
$\cal{L}_{\pi\gamma}^{\rm{raw}}$ are consistent with Eq. ~(A.11) of 
Ref.~\cite{KU}\footnote{ There are misprints in that equation, not present 
in the final result displayed in  Eq.~(3.6) of that work.}, and the result for 
${\cal L}_N^{\rm raw}$ is consistent with that of Ref.~\cite{Ecker}
(to be precise, our ${\cal L}_N^{\rm raw}$ is equal to the final result of 
Ref.~\cite{Ecker}, up to the replacement
$\left[D_{\mu},u^{\mu}\right] \to \frac{i}{2}\chi_{-}-\frac{i}{4}
\left\langle \chi_{-}\right\rangle$).
The result for ${\cal L}_{N\gamma}^{\rm{raw}}$ agrees with Steininger's 
expression~\cite{thesis}, Eq. (3.124) up to several signs
and factors of $i$ [Note e.g. that the expression (3.124) in
Ref.~\cite{thesis} is not hermitean, according to the definition (2.9) in 
that work.].

The ''raw'' divergent Lagrangian, obtained so far, can be consistently used
for renormalization of any process up to the corresponding order. On the other
hand, it is perhaps not the most convenient choice, since it differs from the
standard Lagrangians used in the field. To bring our result to the standard
form, one has to perform the standard procedure, namely use the basis from
Ref.~\cite{GL} for ${\cal L}_{\pi}^{(p^4)}$, and eliminate some of the terms 
by using equations of motion (EOM) for the classical fields.

The EOM are obtained from the linear part of the fluctuation Lagrangian,
\bea \label{eom}
\left[D_{\mu},u^{\mu}\right] &=&\frac{i}{2}\chi_{-}-\frac{i}{4}
\left\langle \chi_{-}\right\rangle 
+4iZF^{2}
\left[ {\cal Q}_{+},{\cal Q}_{-}\right] 
\nonumber\\[2mm]
&+& \frac{i}{4F^{2}}\tau^{a}\bar{N}\left[ \tau^{a},v\cdot u\right]N-
\nonumber\\[2mm]
&-&
\left( \frac{g_{A}}{F^{2}}\tau^{a}\bar{N}\tau^{a}S\cdot
DN+\mathrm{h.c.}\right)\, ,
\nonumber\\[2mm] 
i v\cdot DN &=& -g_{A}S\cdot uN\fs
\eea
Here, we have not included an external source for the nucleon field, because
we are only interested in the $S$-matrix elements.
The proper way of using these EOM is to use appropriate field transformations.
For the mesonic Lagrangian, however, the use of the lowest order EOM, i.e.
replacing the structure $\left[D_{\mu},u^{\mu}\right]$ by the RHS of 
the EOM, is equivalent to the systematic performance of field 
redefinitions~\cite{BCE}. For the baryonic Lagrangian, the required nucleon 
field transformations are explicitly given in~\cite{EM}.

The ~resulting ~divergent ~nucleon Lagrangian corresponds exactly to the heavy 
baryon projections of the Lagrangians ~(\ref{L_p3}) and the pertinent tables 
in the appendix~\ref{app:lagrtables}. For the relativistic Lagrangians at 
$O(p^2),$ $O(p^3),$ ~$ O(e^2p)$, the simple replacements 
$\gamma_{\mu}\gamma_{5}\rightarrow 2 S_{\mu}$, 
$i D_{\mu}\Psi\rightarrow m v_{\mu} N_v$ and
$\sigma_{\mu \nu} \to -2i [S_\mu,S_\nu]$
are all what is needed. Note that in appendix~\ref{app:lagrtables} we do not 
display terms in the Lagrangians ${\cal L}_\pi^{(p^4)}$ and 
${\cal L}_\pi^{(e^2p^2)}$, which do not contain pion fields.
In addition, we drop all terms at $O(e^4)$.

We wish to note that all $\beta$-functions given in the table~\ref{tab:lagr_p4},
coincide with the ones from Refs.~\cite{GL,KU}. The $\beta$-functions from 
table~\ref{tab:LN_p3} are consistent with the results of Ref.~\cite{EM} 
(the latter contains a misprint in $\beta_{11}$, $2/3g_A^5\to -2/3g_A^5$).
Note that in Ref.~\cite{EM}, a different basis in the mesonic sector was used
- consequently, in order to compare, one has to transform the
$\beta$-functions, see below. After this is done, $\beta_i$ from 
table~\ref{tab:LN_p3} are equal to $\beta_i~(i=1\cdots 3)$ and to 
$\beta_{i+1}~(i=4\cdots 23)$ of table~1, Ref.~\cite{EM}. The entries of the 
table~\ref{tab:LN_e2p}, necessary for a coherent result, were, to best of our 
knowledge, not presented yet. As it has been mentioned above, the 
$\beta$-functions for the $O(e^2p)$ Lagrangian, which were presented in 
Ref.~\cite{thesis}, and were used in the numerical fits in Ref.~\cite{Fettes}, 
correspond to the ``raw'' form (\ref{raw-Ngamma}), rather than to the 
Lagrangian which is brought to the ``standard'' form in the mesonic sector 
by using the EOM.

\bigskip

Finally, we comment on the relation between the two most commonly used
choices of operator basis in the ${\cal L}_\pi^{(p^4)}$ Lagrangian.
Some of the $\beta$-functions are changed when one changes the basis
from~\cite{GL} to~\cite{GSS} (the barred quantities belong to the latter
basis) 
\bea\label{GSS-GL}
\bar\beta_{5}&=& \frac{1+5 g_A^2}{24}\, , \qquad \quad
\bar\beta_{18}= 0\, ,\qquad \quad
\nonumber\\[2mm]
\bar\beta_{19}&=& 0\, ,\qquad \quad
\bar\sigma_{8}= \frac{1}{8}\, .
\eea
Furthermore, some of the renormalized LECs do differ in the two bases, namely
\bea\label{d5new}
\bar d_{5}^r(\mu) &=& d_{5}^r(\mu) 
+ \frac{1}{8F^2}\, l_4^r(\mu)\, ,
\nonumber \\
\bar d_{18}&=& d_{18}^r(\mu)
- \frac{g_A}{2F^2}\, l_4^r(\mu)\, , 
\nonumber \\
\bar d_{19}&=& d_{19}^r(\mu)
+ \frac{g_A}{4F^2}\, l_4^r(\mu)\, ,
\nonumber \\
\bar k_{8}^r(\mu) &=& k_{8}^r(\mu) 
+ \frac{1}{2}\,Z l_4^r(\mu)
\fs
\eea

\setcounter{equation}{0}
\section{Integrals}
\label{app:basic}

In this appendix, we present some useful formulae that we have used in
the calculations. Note that all formulae given here correspond to the
case of standard dimensional regularization.

In ~the ~diagrams ~with virtual photons, the following definite integrals over 
the Feynman parameter $x$ are needed [here,  $R=x+(1-x)\,r^2-x(1-x)\,\bar s$ 
is a quadratic polynomial in $x$, and $\bar s\doteq s/m^2$ $\ge (1+r)^2$],

\eq
\int\limits_0^1\! dx\ln R&=&
-2+x_{-}\ln\frac{x_{-}}{1-x_{-}}+
x_{+}\ln\frac{x_{+}}{1-x_{+}}
-i\pi\sigma_0
\nonumber\\[2mm]
&=&
-2+\frac{2\,r\,\ln r}{1+r}
 -i\,\pi\,\sigma_0 +O(\sigma_0)\, ,
\label{R-int_1}
\\[2mm]
\int\limits_0^1\! dx\, \frac{1}{R}&=&\frac{1}{\bar s\,\sigma_0}\,
\left\{
\ln\frac{x_{-}}{1-x_{-}}-\right.
\left.\ln\frac{x_{+}}{1-x_{+}} +2\,i\,\pi
\right\}
\nonumber\\[2mm]
&=&
-\frac{1}{r}
+i\,\pi\,\frac{2}{(1+r)^2\sigma_0}+O(\sigma_0)\, ,
\label{R-int_2}
\\[2mm]
\int\limits_0^1\! dx\, \frac{x}{R}&=&\frac{1}{\bar s\,\sigma_0}\,
\left\{
x_{-}\ln\frac{x_{-}}{1-x_{-}}\right.
\nonumber\\[2mm]
&-&
\left. x_{+}\ln\frac{x_{+}}{1-x_{+}}
+i\,\pi\, (x_{-}+x_{+})
\right\}
\nonumber\\[2mm]
&=&
-\,\frac{1+r+\ln r}{(1+r)^2}
+i\,\pi\,\frac{2\,r}{(1+r)^3\,\sigma_0}+O(\sigma_0)\, ,
\label{R-int_3}\nn
\\[2mm]
\int\limits_0^1\! dx \,\frac{\ln R}{R}&=&\frac{1}{\bar s\,\sigma_0}\,
\left\{
2\,{\rm Li}_2\left(\frac{1-x_{+}}{1-x_{-}}\right)\right.
\left.+2\,{\rm Li_2}\left(\frac{x_{-}}{x_{+}}\right)\right. 
\nonumber\\[2mm]
&+&\left.\frac{4\pi^2}{3}
\right.
\left.+\frac{1}{2}\ln^2\left(\frac{1-x_{+}}{1-x_{-}}\right)
+\frac{1}{2}\ln^2\left(\frac{x_{-}}{x_{+}}\right)\right.
\nonumber\\[2mm]
&+&
\left.
2\,(\ln\bar s+2\ln\sigma_0)\cdot
\left[
\frac{1}{2}\ln\left(\frac{1-x_{+}}{1-x_{-}}\right)\right.\right.
\nn
&+&\left.\left.\frac{1}{2}\ln\left(\frac{x_{-}}{x_{+}}\right)+i\,\pi
\right]
\right\}
\nonumber\\[2mm]
&=&
\frac{\pi^2}{(1+r)^2}\cdot \frac{2}{\sigma_0}-\frac{2}{r}
-\frac{2\,\ln r}{r(1+r)}
\nonumber\\[2mm]
&+&i\,\pi\,\frac{4\,\ln\left[(1+r)\sigma_0\right]}{(1+r)^2\,\sigma_0}
+O(\sigma_0\ln\sigma_0)\, ,
\label{R-int_4}
\en
where
\eq
&&x_{\pm} = \frac{\bar s-1+r^2\pm\lambda^{1/2}(\bar s,1,r^2)}
{2\bar s}\, ,
\nonumber\\[2mm]
&&\sigma_0 = x_{+}-x_{-}=\frac{\lambda^{1/2}(\bar s,1,r^2)}{\bar s}
=\frac{2|{\bf p}|}{\sqrt{s}}\, ,
\nonumber\\[2mm]
&&\lambda(x,y,z)=(x-y-z)^2-4yz\, , \nonumber\\[2mm]
&&{\rm Li}_2(z) = -\int\limits_0^z\! dt\, \frac{\ln(1-t)}{t}\, .
\en

We also recall a useful representation~\cite{tHooft} of the vertex functions 
that occur in the diagrams of Fig.~\ref{fig:strong} $(s_{12})$, $(s_{13})$. These 
are UV and IR finite, so one may put $d=4$ from the beginning.
\eq
(I_3; I_{3v}^\mu)&=& 
\frac{1}{\pi^2i}\,\int d^4k\, \frac{\left(1;\,k^\mu\right)}
{[m_1^2-k^2]}
\nonumber\\[2mm]
&\times&\frac{1}{[m_2^2-(k+p_1)^2]\,[m_3^2-(k+p_1+p_2)^2]}\, ,
\label{3_1}
\\[2mm]
I_{3v}^\mu&=&I_{31}p_1^\mu+I_{32}p_2^\mu\co\nonumber\\[2mm]
I_3&=&\int\limits_0^1\! dx
\left\{
\frac{1}{A_1}\,\ln R_1\right.
\nonumber\\[2mm]
&-&
\left.\frac{(1-\alpha_0)}{A_2}\,\ln R_2
-\frac{\alpha_0}{A_3}\,\ln R_3
\right\}\, ,
\label{3_3}
\\[2mm]
I_{31}&=&\int\limits_0^1\! dx
\left\{
\frac{1}{A_1}
\left[\left(x-1
-\alpha_0\,\frac{R_1}{A_1}\right)\,
\ln R_1+\alpha_0\right]
\right.\nonumber\\[2mm]
&-&\frac{(1-\alpha_0)}{A_2}
\left[\left(x-1-\alpha_0\,\frac{R_2}{A_2}\right)\,
\ln R_2+\alpha_0\,x\right]
\nonumber\\[2mm]
&-&\left.
\frac{\alpha_0}{A_3}
\left[\left(-1-\alpha_0\,\frac{R_3}{A_3}\right)\,
\ln R_3+\alpha_0\,x\right]
\right\}\, ,
\label{3_4}
\\[2mm]
I_{32}&=&\int\limits_0^1\! dx
\left\{
\frac{1}{A_1}
\left[\left(-\frac{R_1}{A_1}\right)\,\right.
\ln R_1+1\right]
\nonumber\\[2mm]
&-&\frac{(1-\alpha_0)}{A_2}
\left[\left(x-1-\frac{R_2}{A_2}\right)\,\ln R_2+x\right]
\nonumber\\[2mm]
&-&
\left.
\frac{\alpha_0}{A_3}
\left[\left(x-1-\frac{R_3}{A_3}\right)\,\ln R_3+x\right]
\right\}\, ,
\label{3_5}
\\[4mm]
R_1&=&x\,m_1^2+(1-x)\,m_2^2-x(1-x)\,p_1^2\, ,
\nonumber\\[2mm]
R_2 &=& x\,m_1^2+(1-x)\,m_3^2-x(1-x)\,p_3^2\, , 
\nonumber\\[2mm]
R_3&=&x\,m_2^2+(1-x)\,m_3^2-x(1-x)\,p_2^2\, ,
\nonumber\\[2mm]
p_3^2&=&(p_1+p_2)^2\, ,
\label{3_6}
\\[4mm]
A_1&=&\lambda^{1/2}(p_1^2,p_2^2,p_3^2)\cdot x+a_1\, ,
\nonumber\\[2mm]
A_2&=&(1-\alpha_0)\,\lambda^{1/2}(p_1^2,p_2^2,p_3^2)\cdot x+a_2\, ,
\nonumber\\[2mm]
A_3&=&-\alpha_0\,\lambda^{1/2}(p_1^2,p_2^2,p_3^2)\cdot x+a_3\, ,
\label{3_7}
\\[4mm]
a_1&=& m_2^2-m_3^2+p_2^2
+\alpha_0\,(m_1^2-m_2^2-p_1^2)\, ,
\nonumber\\[2mm]
a_2&=&a_3=m_2^2-m_3^2-p_2^2
\nonumber\\[2mm]
&+&\alpha_0\,(m_1^2-m_2^2+p_2^2-p_3^2)\, ,
\label{3_8}
\nonumber\\[4mm]
\alpha_0&=& 
\frac{p_1^2+p_2^2-p_3^2+\lambda^{1/2}(p_1^2,p_2^2,p_3^2)}
{2\,p_1^2}\, .
\label{3_9}
\en
The integrals $I_3,I_{31}$ and $I_{32}$ are not changed under scaling of the
arguments of the logarithms by any arbitrary constant value. This property 
allows one to replace $\ln R_i\to\ln R_i/m^2$ in the above expressions.  Note 
that in the threshold amplitude we have $\lambda(p_1^2,p_2^2,p_3^2)=0$. This 
considerably simplifies the calculations.

\setcounter{equation}{0}
\section{Contributions from individual Feynman diagrams}
\label{app:piN-tables}
\begin{figure}[H]
\begin{center}
\resizebox{0.4\textwidth}{!}{\includegraphics{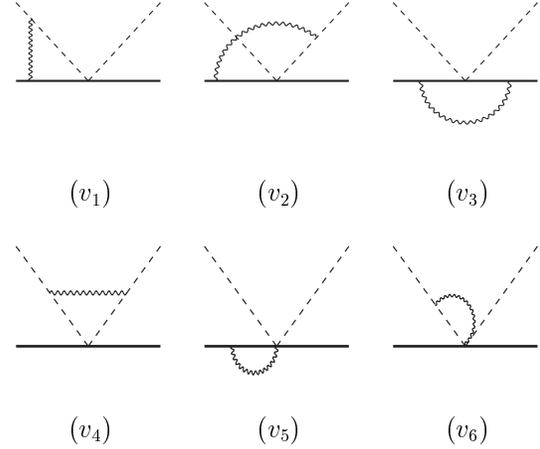}}
\end{center}
\caption{The electromagnetic vector-type diagrams.}
\label{fig:vector}
\end{figure}

\begin{figure*}[htb]
\begin{center}
\resizebox{0.65\textwidth}{!}{\includegraphics[]{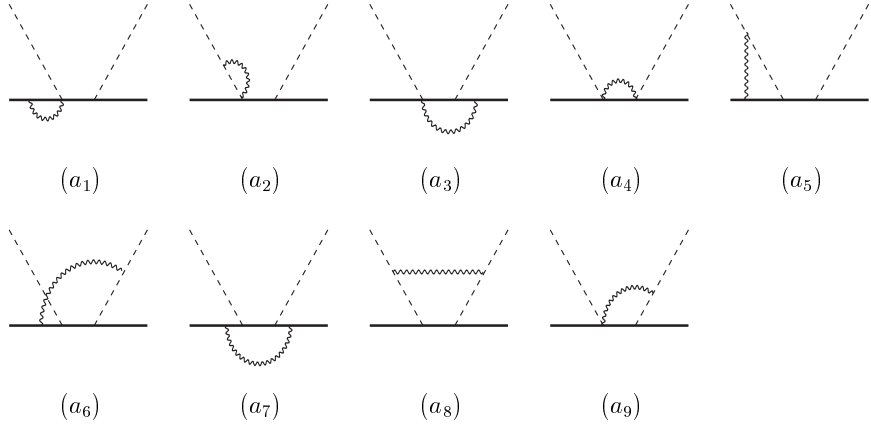}}
\end{center}
\caption{The electromagnetic axial-type diagrams.}
\label{fig:axial}
\end{figure*}

\begin{figure*}[htb]
\begin{center}
\resizebox{0.65\textwidth}{!}{\includegraphics[]{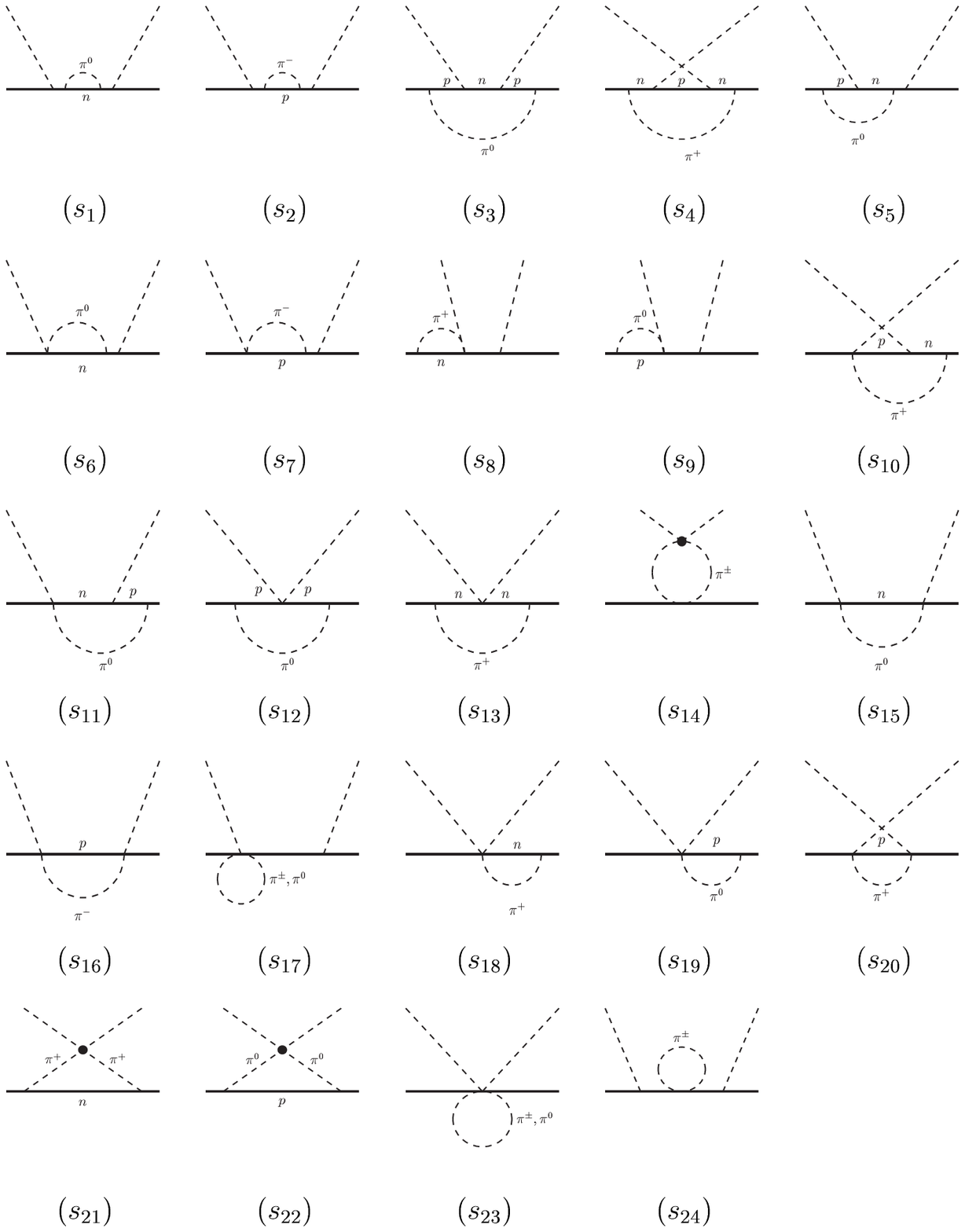}}
\end{center}
\caption{The strong one--loop diagrams.}
\label{fig:strong}
\end{figure*}

In this appendix, we list the non--vanishing contributions to the threshold
amplitude, due to the diagrams displayed in 
Figs.~\ref{fig:vector},\ref{fig:axial} and \ref{fig:strong}.
The notation used is the one in (\ref{TpiN_ini}), (\ref{sums}).
The quantities $C_{\rm UV,IR}$ are defined in Eq.~(\ref{c_uvir}), and
$r=M_\pi/m$.

\begin{table}[H]
\caption{Vector-type electromagnetic diagrams $I_V^i$ in the infrared
  regularization, up to and including $O(r)$.}
\label{tab:V}

\begin{center}
\def\arraystretch{2.0}
\begin{tabular}{|c|c|}
\hline\hline
Fig.~\protect{\ref{fig:vector}} & $I_V^i$ \\
\hline\hline
$(v_1)$ &
$
-\frac{r}{8} 
\left(7\,C_{\rm UV}+8\, C_{\rm IR}+16+30\,\ln{r}\right)$\\
\hline
$(v_2)$ & 
$
\frac{r}{8} 
\left(7\,C_{\rm UV}+8\, C_{\rm IR}+16+30\,\ln{r}\right)$\\
\hline
$(v_3)$ & 
$ \frac{r}{2}\left(C_{\rm UV}-C_{\rm IR}\right)$\\
\hline
$(v_4)$ &
$ -\frac{r}{4}\left(C_{\rm UV}+2\, C_{\rm IR}-4+6\,\ln{r}\right)$\\
\hline
$(v_5)$ & 0 \\
\hline
$(v_6)$ & 
$-\frac{r}{4}\left(-\,3\,C_{\rm UV}+4-6\,\ln{r}\right)$\\
\hline\hline
\end{tabular}
\end{center}
\end{table}

\vspace*{-5.mm}
\begin{table}[H]
\caption{Axial-type electromagnetic diagrams $I_A^i$ in the infrared
  regularization, up to and including $O(r)$.}
\label{tab:A} 
\def\arraystretch{2.0}

\vspace*{.3cm}

\begin{center}
\begin{tabular}{|c|c|}
\hline\hline
Fig.~\ref{fig:axial}  & $I_A^i$ \\
\hline\hline
$(a_4)$  & $-6rC_{\rm UV}+2r-12r(\ln 2+\ln r)$ \\
\hline
$(a_8)$  & $-3rC_{\rm UV}-2r-3r(\pi+2\ln r)$ \\
\hline
$(a_9)$  & $3rC_{\rm UV}-2r+2r(\pi+2\ln 2 +3\ln r)$ \\
\hline\hline
\end{tabular}
\end{center}
\end{table}

\newpage
  
\begin{table}[H]
\caption{Isospin-conserving strong contributions $I_0^i$ and 
iso\-spin--breaking strong contributions $I_\pi^i$ in the infrared
regularization, up to and including $O(r^3)$ and $O(r)$, respectively.}
\label{tab:I0_Ipi}
\def\arraystretch{2.0}

\vspace*{.3cm}

\begin{center}
\begin{tabular}{|c|c|c|}
\hline\hline
Fig.~\ref{fig:strong} & $I_0^i$ &$I_\pi^i$ \\
\hline\hline
$(s_{12})$  & $\frac{g_A^2}{128}\,r^3\,(3\, C_{\rm UV}$  &
 $-\frac{g_A^2}{128}\,r\,(3\, C_{\rm UV}$ \\
&$+2+6\,\ln{r})$ & $+5+6\,\ln{r}) $ \\
\hline
$(s_{13})$  & $-\frac{g_A^2}{64}\,r^3\,(3\, C_{\rm UV}$  & 0 \\
& $ +2+6\,\ln{r})$  & \\
\hline
$(s_{14})$  & $\frac{1}{32}\,r^3\,(C_{\rm UV}+ 2 \ln{r})$ & 0 \\
\hline
$(s_{15})$ & $\frac{1}{128}\,r^3\,(-5\,C_{\rm UV}$  &
 $-\frac{1}{128}\,r\,(3\, C_{\rm UV}$   \\
& $+8-10\,\ln{r}) $& $+11+6\,\ln{r}) $\\
\hline
$(s_{16})$  & $\frac{1}{256}\,r^3\,(-5\,C_{\rm UV}$ & 0 \\
&  $+8-10\,\ln{r}) $  & \\
\hline
$(s_{18})$  & $-\frac{g_A^2}{48}\,\pi\, r^3$ & 0 \\
\hline
$(s_{19})$  & $-\frac{g_A^2}{48}\,\pi\, r^3$ &
$\frac{g_A^2}{32}\,\pi\, r$  \\
\hline
$(s_{20})$  & $-\frac{1}{256}\,r^3\,(-5\,C_{\rm UV}$  & 0 \\
& $+8-10 \ln r) $ & \\
\hline
$(s_{21})$ & $\frac{g_A^2}{96}\,r^3\,(9\, C_{\rm UV}+6$  &
$-\frac{3\,g_A^2}{16} \,\pi\, r $  \\
& $+2\,\pi+18\,\ln{r}) $ & \\
\hline
$(s_{22})$  & $\frac{13\,g_A^2}{192}\,\pi\,r^3$ & 
$-\frac{13\,g_A^2}{128}\, \pi\, r $ \\
\hline
$(s_{23})$ & $-\frac{5}{384}\,r^3\,( C_{\rm UV}+\,2\,\ln{r}) $ &
 $\frac{1}{384}\,r\, (C_{\rm UV}\,$  \\
& & $+1+2 \, \ln{r}) $\\
\hline\hline
\end{tabular}
\end{center}
\end{table}

\newpage

\setcounter{equation}{0}
\section{Figures}
\label{app:figures}

\newpage

\begin{thebibliography}{99}


\bibitem{gm2}
H.N.~Brown {\it et al.}  [Muon g-2 Collaboration],
Phys. Rev. Lett.  \textbf{86}, (2001) 2227
[arXiv:hep-ex/0102017].

\bibitem{DIRAC}
B.~Adeva {\it et al.}, CERN proposal CERN/SPSLC 95-1 (1995). 

\bibitem{PSI-prop}
G.C. Oades at al.,
{\it Measurement of the strong interaction width and shift of the ground 
state of pionic hydrogen}, 
PSI Proposal R-98-01.

\bibitem{PSI}
H.C.~Schroder {\it et al.},
Eur. Phys. J. \textbf{C 21}, (2001) 473.

\bibitem{CGL}
G.~Colangelo, J.~Gasser and H.~Leutwyler, 
Phys. Lett. \textbf{B 488}, (2000) 261 [arXiv:hep-ph/0007112];
Nucl. Phys. \textbf{B 603}, (2001) 125 [arXiv:hep-ph/0103088].

\bibitem{Sigg}
D.~Sigg, A.~Badertscher, P.F.A.~Goudsmit, H.J.~Leisi, and G.C.~Oades,
Nucl. Phys. \textbf{A 609} (1996) 310.

\bibitem{Rasche-piN}
A.~Gashi, E.~Matsinos, G.C.~Oades, G.~Rasche, and W.S.~Woolcock,
arXiv:hep-ph/0009081;
G.~Rasche and W.S.~Woolcock,
Nucl. Phys. \textbf{A 381}, (1982) 405 .

\bibitem{old-pipi} 
V.E.~Lyubovitskij and A.G.~Rusetsky, Phys. Lett. \textbf{B 389}, (1996)
181  [arXiv:hep-ph/9610217];
V.E.~Lyubovitskij, E.Z.~Lipartia, and A.G.~Rusetsky, 
JETP ~Lett. ~\textbf{66}, (1997) 783 [arXiv:hep-ph/9801215];
H.~Jallouli and H.~Sazdjian, Phys. Rev. \textbf{D 58}, (1998) 014011  
[arXiv:hep-ph/9706450];
~P.~Labelle ~and ~K.~Buckley, ~arXiv:hep-ph/9804201;
M.A.~Ivanov, V.E.~Lyubovitskij, E.Z.~Lipartia, and A.G.~Rusetsky, 
Phys. Rev. \textbf{D 58}, (1998) 094024 [arXiv:hep-ph/9805356];
X.~Kong and F.~Ravndal, Phys. Rev. \textbf{D 59}, (1999) 014031;
Phys. Rev. \textbf{D 61}, (2000) 077506 [arXiv:hep-ph/9905539];
B.R.~Holstein, Phys. Rev. \textbf{D 60}, (1999) 114030 
[arXiv:nucl-th/9901041];
D.~Eiras and J.~Soto, Phys. Rev. \textbf{D 61}, (2000) 114027  
[arXiv:hep-ph/9905543];
Phys. Lett. \textbf{B 491}, (2000) 101 [arXiv:hep-ph/0005066];
H.~Sazdjian, Phys. Lett. \textbf{B 490}, (2000) 203 [arXiv:hep-ph/0004226].

\bibitem{Bern1}
A.~Gall, J.~Gasser, V.E.~Lyubovitskij, and A.~Rusetsky, 
Phys. Lett. \textbf{B 462}, (1999) 335 [arXiv:hep-ph/9905309].

\bibitem{Bern2}
J.~Gasser, V.E.~Lyubovitskij, and A.~Rusetsky,
Phys. Lett. \textbf{B 471}, (1999)  244 [arXiv:hep-ph/9910438].

\bibitem{Bern4}
J.~Gasser, V.E.~Lyubovitskij, A.~Rusetsky, and A.~Gall, 
Phys. Rev. \textbf{D 64}, (2001) 016008 [arXiv:hep-ph/0103157]. 

\bibitem{potential-PLB}
E. Lipartia, V.E. Lyubovitskij, and A. Rusetsky,
Phys. Lett. \textbf{B 533}, (2002) 285 [arXiv:hep-ph/0110186].

\bibitem{Bern3}
V.E. Lyubovitskij and A. Rusetsky, 
Phys. Lett. \textbf{B 494}, (2000) 9 [arXiv:hep-ph/0009206].

\bibitem{Hoehler}
G.~H\"{o}hler, in Landolt-B\"{o}rnstein, vol. \textbf{9 b2},
 ed. H.~Schopper (Springer, Berlin, 1983).

\bibitem{sigmaterm}
J.~Gasser, H.~Leutwyler and M.E.~Sainio,
Phys.\ Lett. \textbf{B 253}, (1991) 252 .

\bibitem{GMO}
M.~L.~Goldberger, H.~Miyazawa and R.~Oehme, Phys.\ Rev.
\textbf{99}, (1955) 986 .

\bibitem{Fettes}
N.~Fettes and U.-G.~Mei\ss ner,
Nucl. Phys. \textbf{A 693}, (2001) 693 [arXiv:hep-ph/0101030].

\bibitem{BL1}
T. Becher and H. Leutwyler, Eur. Phys. J. \textbf{C 9}, (1999) 643  
[arXiv:hep-ph/9901384].

\bibitem{Steininger}
U.-G.~Mei\ss ner and S.~Steininger,
Phys. Lett.  \textbf{B 419}, (1998) 403 [arXiv:hep-ph/9709453].

\bibitem{Eiras}
D.~Eiras and J.~Soto,
Phys. Lett. \textbf{ B 491}, (2000) 101 [arXiv:hep-ph/0005066].

\bibitem{Bjorken}
J.D.~Bjorken and S.D.~Drell, Relativistic Quantum Fields,
McGraw-Hill, Inc., 1965.

\bibitem{Yennie} D.R.~Yennie, S.C.~Frautschi, and H. Suura, 
Ann. Phys. \textbf{13}, (1961) 379.

\bibitem{GL}
J.~Gasser and H.~Leutwyler,
Annals Phys.\  \textbf{158}, (1984) 142.

\bibitem{KU}
M.~Knecht and R.~Urech,
Nucl. Phys. \textbf{B 519}, (1998) 329 [arXiv:hep-ph/9709348].

\bibitem{I}
N.~Fettes, U.-G.~Mei\ss ner and S.~Steininger,
Nucl. Phys.  \textbf{A 640}, (1998) 199 [arXiv:hep-ph/9803266].

\bibitem{EM}
G.~Ecker and M.~Mojzis,
Phys. Lett. \textbf{B 365}, (1996)  312  [arXiv:hep-ph/9508204].

\bibitem{Muller}
G.~M\"{u}ller and U.-G.~Mei\ss ner,
Nucl. Phys. \textbf{B 556}, (1999) 265 [arXiv:hep-ph/9903375].

\bibitem{thesis}
S.~Steininger, \textit{``Reelle und virtuelle Photonen 
in chiraler St\"orungstheorie''}, Ph.D. thesis, University of Bonn (1999).

\bibitem{Berezinian}
H.~Neufeld, J.~Gasser and G.~Ecker,
Phys. Lett. \textbf{B 438}, (1998) 106 [arXiv:hep-ph/9806436].

\bibitem{Neufeld}
H.~Neufeld,
Eur. Phys. J. \textbf{C 7}, (1999) 355 [arXiv:hep-ph/9807425].

\bibitem{GSS}
J.~Gasser, M.E.~Sainio and A.~Svarc,
Nucl. Phys. \textbf{B 307}, (1988) 779.

\bibitem{BL2}
T.~Becher and H.~Leutwyler,
JHEP \textbf{0106}, (2001) 017 [arXiv:hep-ph/0103263].

\bibitem{Fettes:2000xg}
N.~Fettes and U.~G.~Mei\ss ner,
Nucl.\ Phys. \textbf{A 676}, (2000) 311 
[arXiv:hep-ph/0002162].

\bibitem{Koch}
R.~Koch, Nucl. Phys. \textbf{A 448}, (1986) 707.

\bibitem{Groom:in}
D.E.~Groom {\it et al.}  [Particle Data Group Collaboration],
Eur. Phys. J. \textbf{C 15}, (2000) 1.

\bibitem{Buttiker}
P.~Buttiker and U.-G.~Mei\ss ner,
Nucl. Phys. \textbf{A 668}, (2000) 97 
[arXiv:hep-ph/9908247].

\bibitem{physrep}
J.~Gasser and H.~Leutwyler,
Phys. Rept.  \textbf{87}, (1982) 77.

\bibitem{Fettes:2000vm}
N.~Fettes and U.-G.~Mei\ss ner,
Phys. Rev. \textbf{C 63} (2001) 045201 
[arXiv:hep-ph/0008181].

\bibitem{Faessler} 
V.E. Lyubovitskij, Th. Gutsche, A. Faessler, and R. Vinh Mau, 
Phys. Lett. \textbf{B 520}, (2001) 204 [arXiv:hep-ph/0108134]; 
Phys. Rev. \textbf{C 65}, (2002) 025202 [arXiv:hep-ph/0109213]. 


\bibitem{photo}
V.~Bernard, N.~Kaiser, J.~Gasser and U.-G.~Mei\ss ner,
Phys. Lett. \textbf{B 268}, (1991) 291;
V.~Bernard, N.~Kaiser and U.-G.~Mei\ss ner,
Z. Phys. \textbf{C 70}, (1996) 483
[arXiv:hep-ph/9411287].

\bibitem{Ball}
R.D. Ball, Phys. Rep. \textbf{182}, (1989) 1.

\bibitem{Ecker}  
G. Ecker, Phys. Lett. \textbf{B 336}, (1994) 508 [arXiv:hep-ph/9402337]. 

\bibitem {BCE}
J. Bijnens, G. Colangelo and G. Ecker, JHEP \textbf{9902}, (1999) 020 
[arXiv:hep-ph/9902437].

\bibitem{tHooft}
G.~t'Hooft and M.J.G.~Veltman,
Nucl. Phys. \textbf{B 153}, (1979) 365.





\end{thebibliography}
\end{document}